\documentclass[useAMS,usenatbib]{mn2e}
\usepackage{subfigure}
\usepackage{graphicx}
\usepackage{amssymb}

\title[The nuclear and extended IR emission of NGC\,2992]{The nuclear and extended infrared emission of the Seyfert 
galaxy NGC\,2992 and the interacting system Arp\,245}
\author[I. Garc\'ia-Bernete et al.]
{\parbox{\textwidth}{I. Garc\'ia-Bernete$^{1,2}$\thanks{E-mail: igarcia@iac.es},
C. Ramos Almeida$^{1,2}$\thanks{Marie Curie Fellow},
J. A. Acosta-Pulido$^{1,2}$,
A. Alonso-Herrero$^{3,4}$,
M. S\'anchez-Portal$^{5}$,
M. Castillo$^{5}$,
M. Pereira-Santaella$^{6}$,
P. Esquej$^{7}$,
O. Gonz\'alez-Mart\'in$^{1,2,8}$,
T. D\'iaz-Santos$^{9,10}$,
P. Roche$^{11}$,
S. Fisher$^{12}$,
M. Povi\'c$^{13}$,
A. M. P\'erez Garc\'ia$^{1,2}$,
I. Valtchanov$^{5}$, 
C. Packham$^{4}$ and N. A. Levenson$^{14}$.
}\vspace{0.4cm}\\
\parbox{\textwidth}{$^{1}$Instituto de Astrof\'isica de Canarias, Calle V\'ia L\'actea, s/n, E-38205, La Laguna, Tenerife, Spain\\
$^{2}$Departamento de Astrof\'isica, Universidad de La Laguna, E-38206, La Laguna, Tenerife, Spain\\
$^{3}$Instituto de F\'isica de Cantabria, CSIC-Universidad de Cantabria, E-39005, Santander, Spain\\
$^{4}$Department of Physics and Astronomy, University of Texas at San Antonio, One UTSA Circle, San Antonio, TX 78249, USA\\
$^{5}$Heschel Science Centre, ESAC, E-28691, Villanueva de la Ca\~nada, Madrid, Spain\\
$^{6}$Centro de Astrobiolog\'ia, CSIC-INTA, E-28850, Torrej\'on de Ardoz, Madrid, Spain\\
$^{7}$Departamento de Astrof\'isica, Facultad de CC. F\'isicas, Universidad Complutense de Madrid, E-28040, Madrid, Spain\\
$^{8}$Centro de Radioastronom\'ia y Astrof\'isica (CRyA-UNAM), 3-72 (Xangari), 8701, Morelia, Mexico\\
$^{9}$Spitzer Science Center, California Institute of Technology, MS 220-6, Pasadena, CA 91125, USA\\
$^{10}$N\'ucleo de Astronom\'ia de la Facultad de Ingenier\'ia, Universidad Diego Portales, Av. Ej\'ercito Libertador 441, Santiago, Chile\\
$^{11}$Department of Physics, University of Oxford, Oxford OX1 3RH, UK\\
$^{12}$Department of Physics, University of Oregon, Eugene, OR 97401, USA\\
$^{13}$Instituto de Astrof\'isica de Andaluc\'ia (IAA-CSIC), E-18008, Granada, Spain \\
$^{14}$Gemini Observatory, Casilla 603, La Serena, Chile
}
}
\begin{document}
\date{}
\pagerange{\pageref{firstpage}--\pageref{lastpage}} \pubyear{2014}
\maketitle
\label{firstpage}
\begin{abstract}

We present subarcsecond resolution infrared (IR) imaging and mid-IR spectroscopic observations of the Seyfert 1.9
galaxy NGC\,2992, obtained with the Gemini North Telescope and the Gran Telescopio
CANARIAS (GTC). The N-band image reveals faint extended emission out to $\sim$3 kpc, and 
the PAH features detected in the GTC/CanariCam 7.5--13~$\mu$m spectrum 
indicate that the bulk of this extended emission is dust heated by star formation.  
We also report arcsecond resolution MIR and far-IR imaging of the interacting system Arp\,245, taken with the 
Spitzer Space Telescope and the Herschel Space Observatory. 
Using these data, we obtain nuclear fluxes using different methods
and find that we can only recover the nuclear fluxes obtained from the subarcsecond data 
at 20-25 $\mu$m, where the AGN emission dominates. We fitted the nuclear IR spectral 
energy distribution of NGC\,2992, including the GTC/CanariCam nuclear spectrum ($\sim$50 pc), with clumpy torus models. 
We then used the best-fitting torus model to decompose the Spitzer/IRS 5--30~$\mu$m spectrum ($\sim$630 pc) in AGN and starburst 
components, using different starburst templates. We find that, whereas at shorter mid-IR wavelengths the starburst component dominates 
(64\% at 6~$\mu$m), the AGN component reaches 90\% at 20~$\mu$m. 
We finally obtained dust masses, temperatures and star formation rates for the different 
components of the Arp\,245 system and find similar values for NGC\,2992 and NGC\,2993. 
These measurements are within those reported for other interacting systems in the first
stages of the interaction.

\end{abstract}


\section{Introduction}
\label{intro}
Active galactic nuclei (AGN) are powered by supermassive black holes (SMBHs), which release enormous quantities of energy 
in the form of radiation or mechanical outflows to the host galaxy interstellar medium. This feedback is fundamental to 
the formation and evolution of the host galaxies \citep{Hopkins10}.
On the other hand, galaxy mergers and interactions or secular processes can generate gas inflows to the nuclear regions of galaxies, 
potentially triggering both AGN and central starbursts (SB; \citealt{Hopkins08}). It has been proposed that the triggering mechanisms might 
depend on AGN luminosity (e.g. \citealt{Ramos12,Treister12}), with high-luminosity AGN (e.g. quasars and powerful radio galaxies) 
being more commonly 
triggered by galaxy interactions and low-to-intermediate luminosity AGN (e.g. Seyferts and low-ionization nuclear emission-line regions; LINERs) 
by disk instabilities, galaxy bars, etc. However, this dependence is not univocal, and examples of both low-luminosity AGN in interacting systems and 
quasars in isolated and morphologically undisturbed galaxy hosts are also found in different galaxy samples \citep{Lipari04,Serra06,Bessiere12}. 

Seyfert galaxies are intermediate-luminosity AGN, characterized by a very bright unresolved nucleus generally hosted by a spiral galaxy \citep{Adams1977}. 
They can be classified as type 1 or type 2 depending on orientation, according to the unified model \citep{Antonucci1993}. This scheme proposes that 
there is dust surrounding the active nucleus distributed in a toroidal geometry, which obscures the central engines of type 2 Seyferts, and allows a 
direct view in the case of type 1. The dusty torus absorbs the intrinsic AGN radiation and, then, reprocesses it to emerge in the infrared (IR), 
peaking in the mid-IR (MIR; $\sim$5-30~$\mu$m).

MIR observations of the nuclear regions of active galaxies allow to study the emission of dust heated by the AGN, 
but also by star-formation (SF) when 
present (e.g. \citealt{Radomski2003,Packham05,Esquej14,Herrero14}). 
Prominent features in the MIR spectrum of Seyfert galaxies are the 9.7~$\mu$m silicate band and the
Polycyclic Aromatic Hydrocarbon (PAH) emission bands, although the latter can be diluted by the bright AGN continuum, and 
therefore they have lower equivalent widths (EWs) than those of non-active star-forming galaxies (see e.g. \citealt{Herrero14, Ramos14}).
The high angular resolution is crucial to correctly separating the nuclear emission from the foreground galaxy emission, as 
the MIR-emitting torus is very small (r $<$ 10 pc; see e.g. \citealt{Tristram09,Burtscher13}).
% Besides, MIR spectroscopy allows to study the most prominent PAH features (6.2, 7.7, 8.6, 11.3 and 17~$\mu$m). 

%The PAHs emission
%mostly originates in photo-dissociation regions where aromatic molecules are heated by the radiation field produced by young massive stars \citep{Roche85}. Thus, PAHs are often used as %indicators
%of the current SFR of galaxies. 
%However, the PAHs are not the best SFR indicator since young massive star are not the only UV source, e.g. AGN emission.

To contribute to the understanding of the relation between nuclear activity, SF, torus properties 
and circumnuclear emission, here we use % high angular resolution 
IR and optical imaging and MIR spectroscopy of the Seyfert galaxy NGC\,2992. This inclined spiral galaxy (b/a=0.31; \citealt{Vaucouleurs91}) 
is located at a luminosity distance of 36.6 Mpc and it is part of the interacting system Arp\,245. This system is formed by NGC\,2992, 
the spiral star-forming galaxy NGC\,2993 \citep{Usui98} and the tidal dwarf galaxy Arp\,245 North (hereafter Arp\,245N; \citealt{duc00}). 
Two bright tidal features connect these three galaxies, suggesting that the system is in an early stage of the interation \citep{duc00}. 
We selected the Arp\,245 system for this IR study, and the galaxy NGC\,2992 in particular, because of the plethora of multiwavelength 
data available in the literature, and because we have new far-IR (FIR) observations from the Herschel Space Observatory as well MIR data 
from CanariCam on the 10.4 m Gran Telescopio CANARIAS (GTC). NGC\,2992 was observed with these telescopes because previous MIR imaging 
data revealed extended emission on nuclear scales, possibly related to the interaction with the other galaxies in the Arp\,245 system. Our 
aim is to study the origin of this IR extended emission by combining the existing and the new IR observations. See Section \ref{obs} for a detailed description of the observations employed here.

NGC\,2992 is classified as a Seyfert 1.9 in the optical, although it has changed its type between Seyfert 1.5 and 2 in the past \citep{Trippe08}. 
It also exhibited huge variations in the X-rays 
(factor of $\sim$20; \citealt{Gilli00}) and in the IR as well \citep{Glass97}. The IR variations were probably caused by a retriggered AGN and by different stages 
of the rebuilding of the accretion disk, with the disk rebuilding estimated to range between 1 and 5 years \citep{Gilli00}. Besides, \citet{Glass97} found flux
variability in the near-IR (NIR), and reported a fading of the source from 1978 to 1996, apart from a strong outburst in 1988. The galaxy shows a thick dust
line at PA$\sim$25-$30^\circ$, measured from north to east, which has been shown to be affecting the emission line profiles in the optical \citep{Colina87}.

%The ultraviolet 
%source heating the dust grains, which remited at IR emission, that can be attributed to cooling dust cloud and it properly explains the IR delayed response to 
%variations in the ultraviolet flux \citep{Clavel89,Barvainis92}.

NGC\,2992 also has evidence for intense SF \citep{Quillen99}, with large-scale outflows observed in H$\alpha$, [O~III]$\lambda$5007 \AA~and soft X-rays, 
driven either by jets or by a SB \citep{Colina87,Colbert98}. In fact, the molecular gas emission in NGC\,2992 could be excited by processes associated with local star 
formation \citep{Quillen99}. The outflow component is distributed in two wide cones \citep{Colina87,Durret87,Colbert96,Allen99,Veilleux01}, with the
geometry of the biconical outflow being such that the southeastern cone is in front of the galaxy disk and the base of the north-western 
outflow is behind it. Apart from the two main kinematic components of the ionized gas (rotation$+$outflow), an additional component is 
required to explain the departure of the ionized gas from the gravitational motion defined from the stars. This component would be related to the AGN, 
and not to the interaction \citep{Garcia01}. The hidden Seyfert nucleus is probably located at the apex outflow origin and at the center of the bulge, 
but it does not seem to coincide with the kinematic center of the disk \citep{Garcia01}.

Although NGC\,2992 has been observed at several wavelengths, up to date there is not any detailed high 
resolution IR analysis of this active galaxy or of the entire system. Here we present a complete IR study of the
interacting system Arp\,245, focusing on NGC\,2992. Section \ref{obs} describes the 
observations and data reduction. The main results, including a compilation of IR fluxes, are presented in Section \ref{results}. We explore
different methods to recover the nuclear emission from low angular resolution data in Section \ref{recover}. We derive relevant
physical parameters of the dust emission in Section \ref{dust}. Finally, in
Section \ref{Discussion} we present the discussion and in Section \ref{Conclusions} we summarize the main conclusions of this work.

Throughout this paper we assumed a cosmology with H$_0$=73 km~s$^{-1}$~Mpc$^{-1}$, $\Omega_m$=0.27, and $\Omega_{\Lambda}$=0.73 for the Arp\,245 system. 
This cosmology provides a luminosity distance of 36.6 Mpc and a spatial scale of 174 pc/$\arcsec$ (from the NASA/IPAC Extragalactic Database; NED).
\section{Observations}
\label{obs}
In this section we describe all the observations analyzed in this work, 
which we divide in subarcsecond and arcsecond resolution data. Subarcsecond data are from
8-10m-class ground-based telescopes and from the Hubble Space Telescope (HST), which allow 
us to resolve the innermost regions of the galaxies. Arcsecond resolution data correspond to 
observations taken with the Spitzer Space Telescope and the Herschel Space Observatory, which 
have lower spatial resolution but higher sensitivity. Details of the observations are summarized in Table \ref{OB}.

\begin{table*}
\centering
\begin{tabular}{cclrrrcc}
\hline
Wavelength &Imaging& Telescope/Instrument  & \multicolumn{2}{c}{Spatial resolution} & Pixel scale&Standard deviation &Date \\
$\lambda_{c}$/$\Delta\lambda$ &Filter/& Imaging & (arcsec)&(pc)& (arcsec/pixel) &$\sigma$ &(UT) \\
($\mu$m)	&Band 	&   & & & &(10$^{-2}$)& \\
\hline
0.6/0.15 		&F606W	& HST/WFPC2 &0.095&17&0.046&1.3&1994 Oct 18\\%500.00\\
2.07/0.6		& F205W	& HST/NICMOS2 &0.107&19&0.075&12.9&1998 Oct 15\\% 383.75\\
3.6/0.75		& Ch1	& SPITZER/IRAC&1.85&322&0.6&0.01&2004 Dec 21\\ %10.40	\\
4.5/1.02 		& Ch2& SPITZER/IRAC&1.77&308&0.6&0.01&2004 Dec 21\\ %10.40	 \\
5.8/1.43		& Ch3& SPITZER/IRAC&2.15&374&0.6&0.08&2004 Dec 21\\%10.40	\\
8/2.91			& Ch4& SPITZER/IRAC&2.79&485&0.6&0.05&2004 Dec 21	 \\
11.2/2.4		&N'	& GEMINI/MICHELLE&0.32&56&0.1005&7.6&2006 May 12	\\%150.0 \\
18.1/1.9 		&Qa& GEMINI/MICHELLE&0.53&92&0.1005&68.4&2006 May 12\\%150.00	 \\
24/4.7			&Ch1& SPITZER/MIPS&6.06&1054& 1.225&0.2&2008 Jun 23 \\%2.62	\\
70/10.6			&Blue& HERSCHEL/PACS&5.25&914& 1.4&8.9&2011 May 16 \\ %1007.00	\\
100/17			&Green& HERSCHEL/PACS&6.75&1175& 1.7&12.3&2011 May 16\\ %1007.00	\\
160/30.2		&Red& HERSCHEL/PACS&10.80&1879& 2.85&13.1&2011 May 16\\ %1007.00??	\\
250/75.76		&PSW& HERSCHEL/SPIRE&17.63&3068&6&48.3&2010 Jun 21\\ %445.00??	\\
350/102.94		&PMW& HERSCHEL/SPIRE&24.49&4261&10&94.3&2010 Jun 21\\ %445.00??	\\
500/200			&PLW&  HERSCHEL/SPIRE&34.66&6031&14&52.4&2010 Jun 21\\ %445.00??	\\
%8.7		&2014 Feb 13	& GTC&CANARICAM& Si-2&695\\%695.00	 \\
\hline
Wavelength & Spectroscopy& Telescope/Instrument& \multicolumn{2}{c}{Spectral resolution}& Pixel scale&Exposure&Date \\
($\mu$m) &Slit width& Spectroscopy  &  \multicolumn{2}{c}{$\lambda$/$\Delta\lambda$} & (arcsec/pixel)&time &(UT) \\
	 &(arcsec) & & & & &(s)\\
\hline
7.5-13& 0.52 & GTC/CANARICAM&\multicolumn{2}{c}{$\sim$175}&0.0798&943&2014 Feb 13\\%943.22\\
8-13& 0.4 & GEMINI/MICHELLE&\multicolumn{2}{c}{$\sim$200}&0.183&1200&2007 Mar 25\\%.00\\
5.2-14.5& 3.6-3.7 & SPITZER/IRS~SL&\multicolumn{2}{c}{60-127}&1.8&14&2005 Dec 10	\\ %14.68\\
14-38& 10.5-10.7 & SPITZER/IRS~LL&\multicolumn{2}{c}{57-126}&5.1&6&2005 Dec 10	\\ %14.68\\
%7.5-13& 0.52 & GTC/CANARICAM&0.27&47&0.0798&$\sim$175&2014 Feb 13\\%943.22\\
%8-13& 0.4 & GEMINI/MICHELLE&0.38&66&0.183&$\sim$200&2007 Mar 25\\%.00\\
%5.2-14.5& 3.6-3.7 & SPITZER/IRS~SL&$\sim$3&$\sim$522&1.8&60-127&2005 Dec 10	\\ %14.68\\
%14-38& 10.5-10.7 & SPITZER/IRS~LL&$\sim$3&$\sim$522&5.1&57-126&2005 Dec 10	\\ %14.68\\
\hline
\end{tabular}						 
\caption{Summary of the imaging and spectroscopic observations. The $\sigma$ corresponds to the standard 
deviation of the sky background in mJy/pixel units. Note that the PACS observations were done using 
the normal/cross scan pattern, whereas the small map SPIRE observations were done using the nominal/cross scan pattern.}
\label{OB}
\end{table*}

\subsection{Subarcsecond resolution data}
\subsubsection{MIR Gemini/MICHELLE observations}
\label{michelle}

\begin{figure*}
\centering
\includegraphics[width=5.9cm,angle=90]{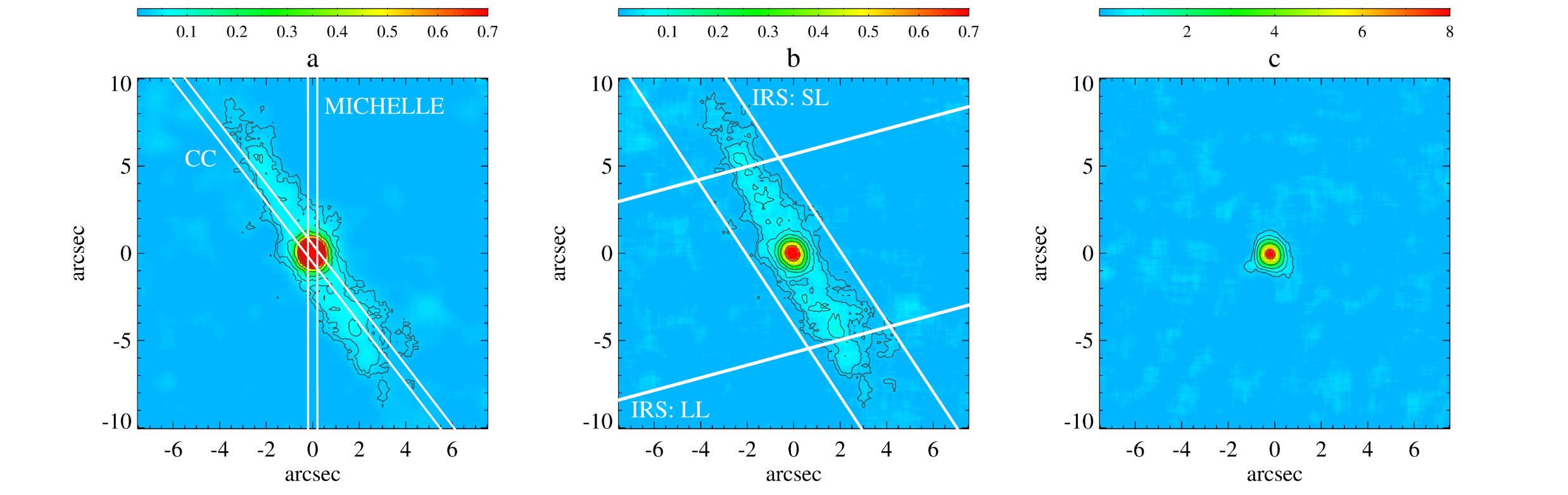}
\caption{Gemini/MICHELLE images of NGC\,2992. (a) MICHELLE~11.2~$\mu$m image with the CanariCam and MICHELLE slits overlaid. 
(b) PSF-subtracted MICHELLE 11.2~$\mu$m image with the Spitzer/IRS slits overlaid (see Section \ref{spitzer_obs}). 
(c) MICHELLE 18.1~$\mu$m image. All images are smoothed (box of 6 pixels) and have their own contours overlaid (in black). 
Colour bars correspond to fluxes in mJy units. North is up, and east to the left.}
\label{Np_smooth}
\end{figure*}

Two images were taken in the N' ($\lambda_c$=11.2~$\mu$m) and Qa ($\lambda_c$=18.1~$\mu$m) filters using the instrument MICHELLE 
\citep{Glasse97} on the 8.1 m Gemini-North Telescope. 
MICHELLE is a MIR (7-26~$\mu$m) imager and spectrograph, which
uses a Si:As detector, covering a field-of-view (FOV) of
32\arcsec$\times$24\arcsec~on the sky. Its pixel scale is 0.1005\arcsec. The standard MIR chopping-nodding technique was used to
remove the time-variable sky background and the thermal emission from the telescope.  The chopping and nodding throws were 15\arcsec,
optimal for the galaxy size (minor axis$\sim$2\arcsec) and perpendicular to the semi-major axis of the galaxy\footnote{The chopping 
throw was chosen according to the galaxy size in the Gemini/MICHELLE image. We note that both the resolution and sensitivity of the 
Spitzer Space Telescope and Gemini are completely different, and thus, the large-scale IR emission that we detect in the Spitzer 
images is completely absent in the ground-based image and it does not affect the latter.}. The on-source integration times were 150 s for both galaxy images. Besides,
images of a point spread function (PSF) standard star were obtained immediately after the science target in the N' and Qa filters 
for accurately sampling the image quality, and to allow flux calibration. 
We measured angular resolutions of 0.32\arcsec~and 0.53\arcsec~from the full width at half-maximun (FWHM) of these PSF standard stars.
The fully reduced images of NGC\,2992, taken from \citet{Ramos09}, are presented in Fig. \ref{Np_smooth}.
The N-band image shows a morphology consisting of a dominant point source and faint extended emission along PA$\sim30^\circ$, whereas 
the Q-band image shows unresolved emission only. See Section \ref{large} for more details.

In addition, a MIR spectrum covering the spectral range 8-13~$\mu$m was obtained with the low resolution (R=$\lambda$/$\Delta\lambda\sim$200) MICHELLE N-band 
grating. A slit of $\sim$0.4\arcsec~width was used, oriented at $PA=0^\circ$, and the on-source integration time was 1200 s.
We used the reduced and flux calibrated spectrum from \citet{Colling11}, also presented in \citet{Esquej14}.

%Data are part of Chris Packham observations (Observation ID: GN-2006-Q11).

%The imaging reduction was carried out with the Gemini {\textit{IRAF}}
%packages, particularly with the {\textit{MIDIR}} \citep{tody86} reduction task. The Gemini {\textit{IRAF}} packages includes sky subtraction,
%stacks of individual images and rejection of bad images. The flux calibration was carried out using the Phot {\textit{IRAF}} task to 
%making a relation between counts and real flux. The spectral reduction was carried out with {\textit{RedCan}} in the same way that the CC data.

%We remark that although the Qa image is like a punctual 
%source image, the N' image has extended emission that shown in Fig. \ref{Np_smooth}. For more details see section 3.1.1.

\subsubsection{MIR Gran Telescopio CANARIAS/CanariCam observations}
\label{canaricam}

We obtained a N-band spectrum (7.5-13~$\mu$m) of NGC\,2992, using the low spectral resolution (R$\sim$175) 
grating available in the instrument 
CanariCam (CC; \citealt{Telesco03}), on the 10.4m GTC. 
CC is a MIR (7.5--25~$\mu$m) imager with spectroscopic, 
coronagraphic and polarimetric capabilities and uses a Si:As detector, which covers a FOV of
26\arcsec$\times$19\arcsec on the sky and it has a pixel scale of 0.0798\arcsec (hereafter 0.08\arcsec). The slit, of width $\sim$0.52\arcsec, was 
oriented at PA$=30^\circ$, following the faint extended emission revealed by the 
Gemini/MICHELLE imaging data (see Fig. \ref{Np_smooth}). The total on-source integration time was 943 s and
the standard MIR chopping-nodding technique was used, with chop and nod throws of 15\arcsec.

The data were taken as part of an ESO/GTC large programme (182.B-2005; PI: Alonso-Herrero), aimed to conduct a MIR survey of nearby AGN by
exploiting the unique capabilities of CC on the GTC (see \citealt{Herrero13,Herrero14} for further details).  The data reduction 
was carried out with the {{\textit{RedCan}} pipeline \citep{Gonzalez-Martin13}, which 
performs sky subtraction, stacking of individual observations, rejection of bad frames, wavelength calibration, trace determination and spectral 
extraction. We extracted a nuclear spectrum as a point source, and another spectrum as an extended source in an aperture radius of 
5.2\arcsec~($\sim$905~pc), which we then use to characterize the extended emission (see Section \ref{spectroscopy}).
% two spectra at both sides of the nucleus with an offset of 0.6\arcsec~in each direction, in this case as extended source. 
Note that
in the case of point source extraction, {\textit{RedCan}} uses an aperture that increases with wavelength to take into account
the decreasing angular resolution, and it also performs a correction to account for slit losses. For the extended source extraction, 
a fixed 5.2\arcsec~aperture and no slit-loss corrections were 
applied (see \citealt{Gonzalez-Martin13} for further details on CC data reduction).

\subsubsection{Optical and NIR HST observations}
\label{hubble}

We downloaded the fully reduced optical and NIR imaging data of NGC\,2992 from the ESA Hubble Legacy 
Archive\footnote{http://archives.esac.esa.int/hst/}. 
The optical image, shown in the left 
panel of Fig.~\ref{hst_vs_gemini}, was observed with the F606W filter ($\lambda_c$=5975~\AA) using the 
Wide Field Planetary Camera 2 (WFPC2), which has a FOV of 2.7\arcmin$\times$2.7\arcmin~on the sky and a pixel scale of 0.046\arcsec. 
The NIR image (see right panel of Fig.~\ref{hst_vs_gemini}) was observed with
the F205W filter ($\lambda_c$=2.07~$\mu$m) using the Near Infrared Camera and Multi-Object
Spectrometer (NICMOS), which has a FOV of 19.2\arcsec$\times$19.2\arcsec~on the sky and a pixel scale of 0.075\arcsec (NIC2 camera).
The downloaded NICMOS and WFPC2 images were reduced using the {\textit{NICRED}} package \citep{McLeod97} and the Space Telescope 
Science Analysis System ({\textit{STSDAS}}) within {\textit{IRAF}}\footnote{IRAF 
is distributed by the National Optical Astronomy Observatory, which is operated by the Association of Universities for Research in 
Astronomy (AURA) under cooperative agreement with the National Science Foundation.}.

\begin{figure*}
\centering
\includegraphics[width=10cm, angle=90]{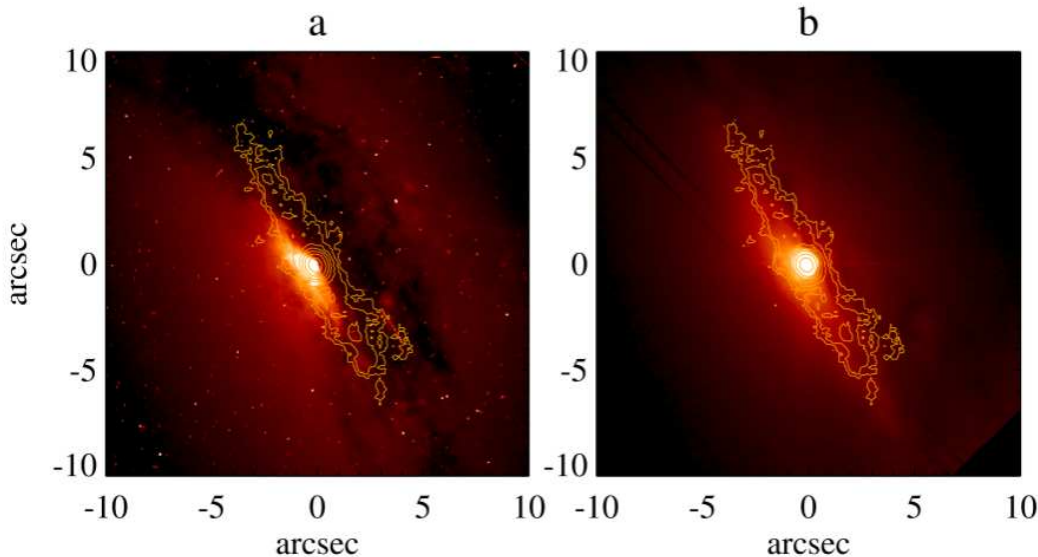}
\caption{HST/WFPC2 optical image of NGC\,2992 in the F606W filter (left panel), and HST/NICMOS NIR image in the F205W filter 
(right panel). Orange contours correspond to the PSF-subtracted Gemini/MICHELLE 11.2~$\mu$m image (panel b in Fig. 1). 
North is up, and east to the left. }
\label{hst_vs_gemini}
\end{figure*}

The optical and NIR data were taken as part of the Hubble programs P5479 (cycle:4, PI: M. Malkan) and P7869 (cycle:7, PI: A. Quillen), respectively. 
We refer the reader to \citet{Malkan98} and \citet{Quillen99} for further details on these HST observations.

%We used data reduced from the archive, which were reduced using the software of The Space Telescope Science Institute (STScI), the Space %Telescope Science Analysis System ({\textit{STSDAS}}), 
%for reducing and analyzing HST data. This software works within {\textit{IRAF}}.

\subsection{Arcsecond resolution data}
\subsubsection{MIR Spitzer Space Telescope observations}
\label{spitzer_obs}

We donwloaded imaging data of the Arp 245 system from the Spitzer Heritage Archive (SHA), taken 
with the instruments Infrared Array Camera (IRAC; \citealt{Fazio04}) and Multiband Imaging
Photometer for Spitzer (MIPS; \citealt{Rieke04}). The IRAC FOV is 5.2\arcmin$\times$5.2\arcmin~on the sky 
and its pixel scale is 1.2\arcsec, whereas the MIPS
FOV is 5.4\arcmin$\times$5.4\arcmin~on the sky and its pixel scale is 2.45\arcsec. 

In addition, a low resolution MIR spectrum of NGC\,2992 was retrieved from the Cornell Atlas 
of Spitzer/IRS Source (CASSIS v4; \citealt{Lebouteiller11}). The spectrum
 was obtained using the InfraRed Spectrograph (IRS; \citealt{Houck04}). The observation was 
 made in staring mode using the two low-resolution (R$\sim$60-120) IRS modules: the short-low 
 (SL; 5.2-14.5~$\mu$m) and the long-low (LL; 14-38~$\mu$m). The slits were oriented as shown in 
 Fig. \ref{Np_smooth} and their widths are 3.6\arcsec~and 10.5\arcsec~for the SL and LL modules respectively. 

The IRAC imaging data (3.6, 4.5, 5.8 and 8.0~$\mu$m) and the IRS spectrum were taken as part of the Spitzer program P96 
(PI: J. R. Houck), and the MIPS imaging data (24~$\mu$m) under program P40936 (PI: G. Rieke). 
In the case of the IRAC images, we just downloaded the mosaicked data from the Spitzer archive, which have a pixel scale 
of 0.6\arcsec. On the other hand, the MIPS image was reprocessed using the MOsaicking and Point source EXtraction ({\textit{MOPEX}}) software. 
Background gradients were removed by self-calibrating 
the data (see Section 8.1 of the \citealt{MIPS11} for details) and the resulting mosaics were resampled to a pixel size of 1.225\arcsec.

%The BCD processing includes corrections for the instrumental response, flagging of cosmic rays and saturated
%pixel, dark and flat-fielding corrections, and flux calibration based on standard star. The mosaic images are in 
%MJy$/$sr units, and we used different conversion factors, depending on the pixel size, to convert their to mJy/pixel. And 
We downloaded the IRS spectrum from the CASSIS database. The spectrum was reduced 
with the CASSIS software, using the optimal extraction to get the best signal-to-noise ratio. We only needed to apply a small
offset to stitch together the different modules, taking the shorter wavelength module (SL2; 5.2-7.6~$\mu$m) as the basis, which 
has associated a slit width of 3.6\arcsec~($\sim$630 pc). The IRS spectrum is shown in left panel of Fig. \ref{spec_ngc2992} 
(black dashed line) and it was also presented in \citet{Esquej14}.

\subsubsection{FIR Herschel Space Observatory observations}
\label{herschel}

FIR imaging data of the Arp 245 system were obtained with the Photodetector Array Camera and Spectrometer (PACS; \citealt{Poglistch10}) and
the Spectral and Photometric Imaging REceiver (SPIRE; \citealt{Griffin10}) on-board of the 
Herschel Space Observatory \citep{Pilbratt10}.  
The data are part of the guaranteed time proposal ``Herschel imaging photometry of nearby Seyferts galaxies: Testing the coexistence 
of AGN and SB activity and the nature of the dusty torus'' (PI: M. S\'anchez-Portal). 

The PACS instrument 
has a FOV of 1.75\arcmin$\times$3.5\arcmin~on the sky and three different bands (70, 100 and 160~$\mu$m),
with beam sizes of 5.6, 6.8 and 11.3 arcsec FWHM, respectively.
The PACS observations were carried out using the mini-map mode, consisting of two concatenated 3\arcmin~scan line maps, at 70\degr~and 
110\degr~(in array coordinates). This results in a map with a highly homogeneous exposure within the central 1\arcmin~area. 
The SPIRE instrument
has a FOV of 4\arcmin$\times$8\arcmin~on the sky and three different bands (250, 350 and 500~$\mu$m),
whose beam sizes are 18.1, 25.2 and 36.9 arcsec FWHM, respectively. These three available bands were observed simultaneously using the 
small map mode, whose area for scientific use is around 5\arcmin$\times$5\arcmin.

%The SPIRE instrument
%has a FOV of 4\arcmin x8\arcmin~on the sky and three different bands (250, 350 and 500~$\mu$m),
%whose beam sizes are 18.1, 25.2 and 36.9 arcsec FWHM, respectively. 
%The data cover the spectral range 70-500~$\mu$m in six different bands. 

%The {\bf{PACS data}} reduction was carried out with the Herschel Interactive Processing Environment 
%({\textit{HIPE}}; \citealt{Ott10}) v8.0.1 and {\textit{Scanamorphos}} \citep{Roussel12} v15. For the PACS instrument, we used {\textit{HIPE}} and Calibration Database
%V32 to build Level 1 products. These tools perform the data reduction, which
%included detecting and flagging bad pixels, converting the analog to digital units readings to flux units (Jy/pixel) and adding the pointing information. We
%did not attempt to perform deglitching at this stage to prevent the bright AGN nucleus to be affected by the multi-resolution median transform deglitching process. The final maps were built
%form the Level 1 products using {\textit{Scanamorphos}}, which performs a baseline subtraction, correction of the striping effect due to the scan process, removal of global and individual pixel drifts, and finally the map
%assembly using all the nominal and coss-direction scans. 

The PACS data processing was carried out by means of two tools:  the Herschel Interactive Processing Environment 
({\textit{HIPE}}; \citealt{Ott10}) and {\textit{Scanamorphos}} \citep{Roussel12}. In order to build the Level 1 products 
we used {\textit{HIPE}} v8.0.1 with the PACS calibration database V32. This Level 1 processing included detecting and 
flagging bad pixels, converting  the analogue to digital units readings to flux units (Jy/pixel) and adding the pointing 
information.  We did not attempt to perform deglitching at this stage to prevent the bright AGN nucleus to be affected by 
the multi-resolution median transform deglitching process. The final maps were built from the Level 1 products using 
{\textit{Scanamorphos}} v15, which performs a baseline subtraction, correction of the striping effect due to the scan 
process, removal of global and individual pixel drifts, and finally the map assembly using all the nominal and cross-direction 
scans.

For the SPIRE data processing we built the Level 1 products with {\textit{HIPE}} v8.0.1 and the SPIRE calibration 
database v8.1. The Level 1 processing included detection of thermistor jumps in the time line, frame deglitching,
low-pass filter correction, conversion of readings to flux units (Jy/beam), temperature drift and bolometric time 
response corrections, and addition of pointing information.  The final maps were built from the Level 1 using the 
Na\"ive Mapper functionality integrated in HIPE v8.0.1. This mapping  strategy simply projects the integrated power 
seen by each bolometer onto the nearest sky map pixel. Once all the detector signals have been mapped, the flux density
map and the standard deviations are calculated.

%For SPIRE we used the standard (small) map script and Calibration Database v8.1. The processing included detection of thermistor jumps in the time line, 
%frame deglitching, low-pass filter correction, conversion of readings to flux units (Jy/beam), temperature drift and bolometric time response corrections, and addition of pointing information. We built the final maps
%using the ``Naive'' scan mapper task. %Finally we changed the flux units (Jy/beam) to (Jy/pixel) for each beam size.

\section{IR EMISSION OF THE SYSTEM}
\label{results}

\subsection{Nuclear region of NGC\,2992}%emission: The Innermost 100 pc}
\label{nuclear}

In this section we study in detail the properties of the inner 100 pc of the galaxy NGC\,2992 as well as the surrounding faint extended emission.

\subsubsection{Imaging}
\label{imaging}
%\textcolor{red}{Nota Jose: Hablar antes de la emission nuclear y despues de la emision extensa....pero hicimos esto por remarcar la importancia de la emision extensa que no estaba publicada hasta ahora, no?}

In Fig. \ref{Np_smooth} we show the high angular resolution MIR Gemini/MICHELLE images of NGC\,2992. 
The 11.2~$\mu$m image reveals faint emission along PA$\sim$30$^\circ$ and extending out to $\sim$3 kpc (see more details in Section \ref{large_2992}). 
This emission is coincident
with the extended emission shown in the NIR HST image (right panel of Fig. \ref{hst_vs_gemini}). 
On the other hand, the optical HST image shows a thick dust lane that partly obscures the nucleus and whose orientation coincides with that 
of the extended emission (30$^\circ$; left panel of Fig. \ref{hst_vs_gemini}). For the image registration, we first used the galaxy nucleus 
in the MICHELLE and NICMOS images, and then we used different stars in the FOV that the NICMOS and WFPC2 images have in common for centering 
the optical image. We note that the galaxy nucleus in the HST/NICMOS image is saturated. Finally, we measured a surface brightness of 
4.84$\pm$0.07 mJy/arcsec$^2$ for the extended emission in the 11.2~$\mu$m image with the {\textit{PHOT}} task of {\textit{IRAF}}.
We used an aperture of 1\arcsec~diameter in two different positions at both sides of the nucleus and we averaged the two measurements.

In addition to the extended emission seen in the 11.2~$\mu$m image there is an unresolved nuclear component, which is also present in 
the Q-band image. \citet{Ramos09} estimated MIR nuclear fluxes of 175 and 521 mJy in the N- and Q-bands, respectively, by
subtracting the PSF standard stars, observed in each filter before or after the science observations, from the galaxy profiles. In 
the central panel of Fig. \ref{Np_smooth} we show the PSF-subtracted N-band image of the galaxy, with the PSF scaled at 90\%.

\subsubsection{Spectroscopy}
\label{spectroscopy}
In the left panel of Fig. \ref{spec_ngc2992} we show the GTC/CC and Gemini/MICHELLE 8-13~$\mu$m nuclear spectra of NGC\,2992, 
both extracted as a point source.
The GTC/CC spectrum
has a spatial resolution of 0.27\arcsec~and the Gemini/MICHELLE spectrum of 0.38\arcsec, which correspond to physical scales of $\sim$47 pc 
and $\sim$66 pc, respectively. Both angular resolutions were determined from the FWHM of the PSF star acquisition images.

\begin{figure*}
\centering
\par{
\includegraphics[width=5.85cm,angle=90]{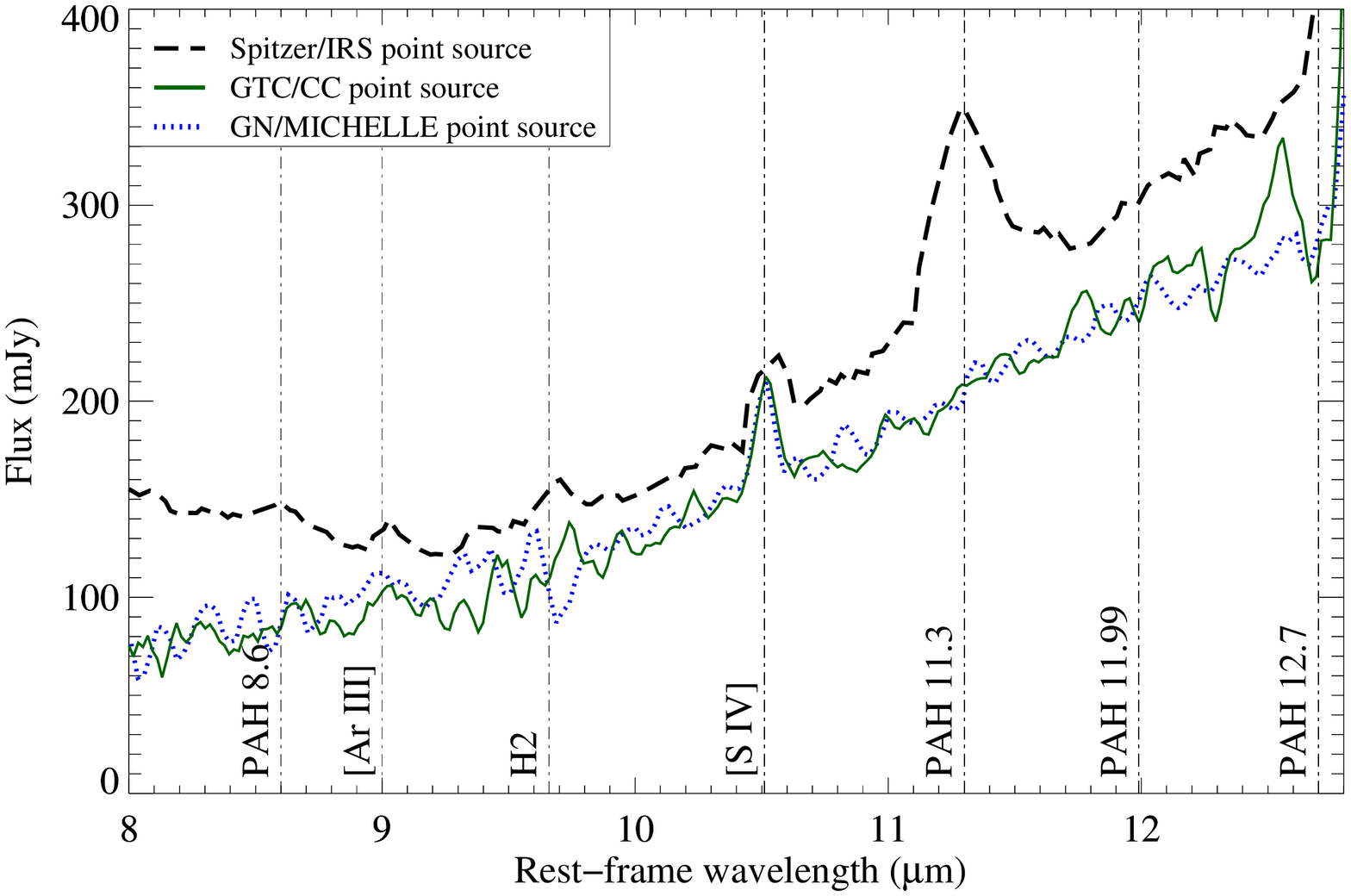}
\includegraphics[width=5.85cm,angle=90]{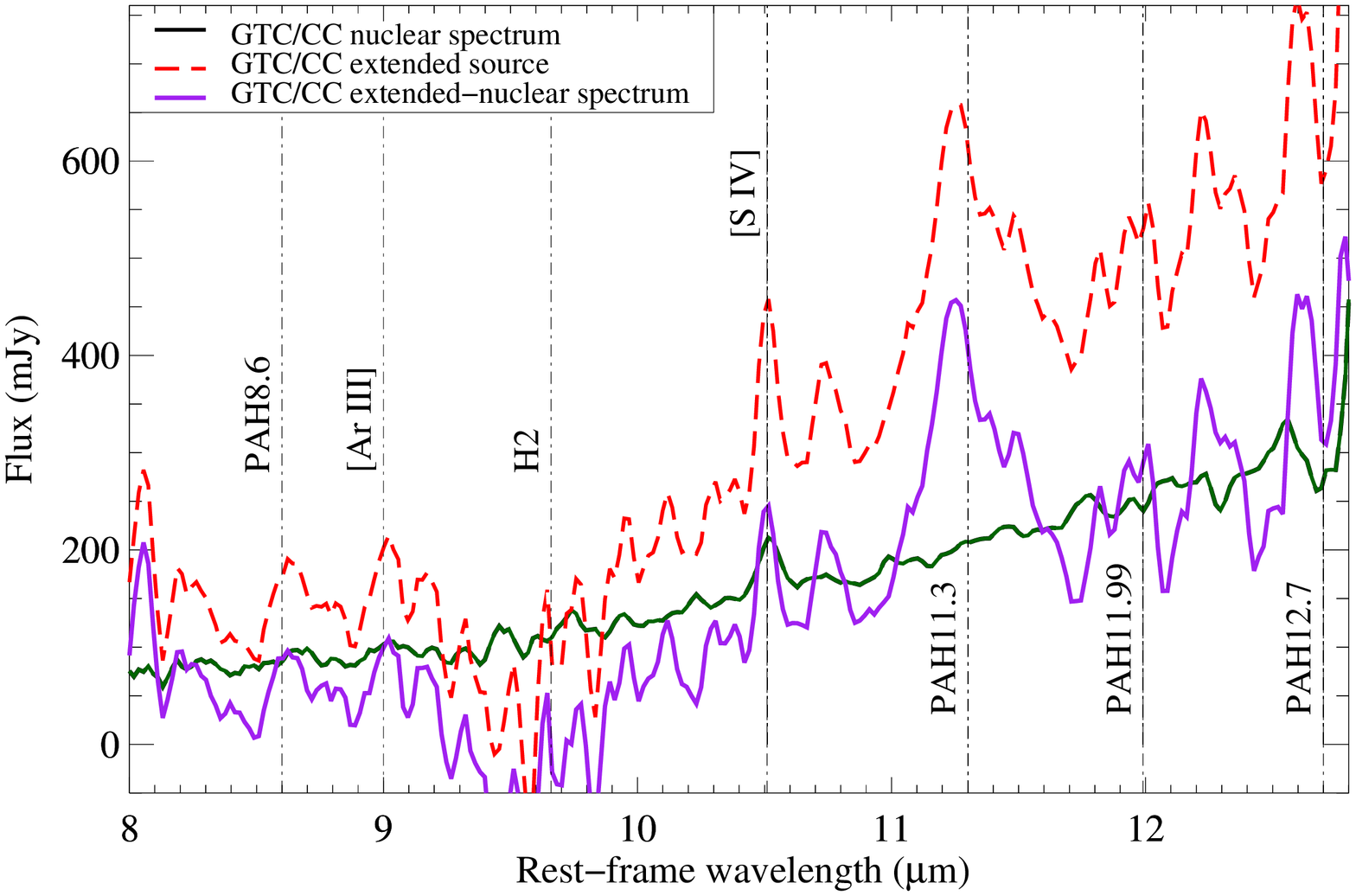}\par} 
\caption{Left panel: Spitzer/IRS rest-frame spectrum ($\sim$630 pc) of NGC\,2992 (dashed black line) and GTC/CC and Gemini/MICHELLE 
rest-frame nuclear spectra ($\sim$47 and $\sim$66 pc) of NGC\,2992 (solid green and dotted blue lines respectively), extracted as a 
point source. Right panel: GTC/CC spectrum of NGC\,2992 extracted as an extended source in an aperture radius of 
5.2\arcsec~$\sim$905~pc (dashed red line), GTC/CC 
nuclear spectrum (as in left panel; solid green line) and spectrum of the extended emission (solid purple line), obtained by subtracting
the nuclear spectrum from the one extracted in the 5.2\arcsec~aperture radius. In both 
panels, the vertical dotted lines indicate the position of typical star-forming regions/AGN emission lines/bands. All spectra 
have been smoothed (2 pixel box).}
\label{spec_ngc2992}
\end{figure*}

%(more than nine times smaller).
Despite the different slit orientations (see Fig. \ref{Np_smooth}) and the time difference between the observations ($\sim$7 years), 
both spectra are practically identical. In spite of the IR variability reported
by \citet{Glass97}, we do not see any difference either in flux or shape. These spectra do not show PAH features and they exhibit
[S IV]$\lambda$10.5~$\mu$m emission, which can originate in the Narrow-Line Region (NLR) and it is usually considered an AGN tracer 
\citep{dasyra11}. However, 
this emission line can also be produced in star forming regions, as shown by \citet{pereira10}, due to its relatively 
low ionization potential (35 eV). For comparison, in the left panel of Fig.~\ref{spec_ngc2992} we show the Spitzer/IRS 
spectrum in the same spectral range as the GTC/CC and Gemini/MICHELLE spectra, which
has a spatial resolution of 3.6\arcsec, that corresponds to a physical scale of $\sim$630 pc. Unlike the nuclear spectra, 
the IRS spectrum shows 8.6 and 11.3~$\mu$m PAH bands, indicative of the presence of SF on the scales probed by Spitzer.

In the right panel of Fig. \ref{spec_ngc2992} we show the nuclear GTC/CC spectrum, extracted as an extended source in an 
aperture radius of 5.2\arcsec~($\sim$905 pc), the GTC/CC nuclear spectrum extracted as point source and the spectrum of the extended emission.
The latter was obtained by subtracting the nuclear spectrum from the one extracted in the large aperture, in order to get
rid of the AGN contribution.
%The nuclear spectrum extracted as an extended source has a big contribution of the AGN, therefore, we subtracted the 
%nuclear spectrum extracted as a point source to minimize the AGN contimination, and thus we can obtain the spectrum of
%the extended emission. 
We chose this large aperture to increase the signal-to-noise of the extended emission 
spectrum. 
The spectra of the extended emission, before and after subtracting the AGN contribution, clearly show
11.3 $\mu$m PAH feature and the [S~IV]$\lambda$10.5~$\mu$m emission line, exactly as the Spitzer/IRS spectrum
on scales of $\sim$630 pc. Thus, the faint extended emission that we detect in the Gemini/MICHELLE N-band image of the galaxy 
is, as least in part, due to SF. On the other hand, if we compare these spectra with the GTC/CC and Gemini/MICHELLE 
nuclear spectra shown in the left panel of Fig. \ref{spec_ngc2992}, we can conclude that either the PAH features 
have been destroyed in the inner $\sim$50 pc of NGC\,2992, or are diluted by the strong AGN 
contiuum (see \citealt{Herrero14} and \citealt{Ramos14}).

%as shows the Spitzer/IRS spectrum at bigger scales. This indicates that, at least in the case of NGC\,2992, this extended emission would be 
%tracing SF, whereas in the innermost $\sim$60 pc the SF activity is decreased or is diluted 
%by the AGN.}

\subsubsection{Nuclear SED modelling with clumpy torus models}
\label{bayesclumpy}

Recent studies assumed a clumpy distribution of dust surrounding AGN to explain the properties of 
the nuclear IR SED of Seyfert galaxies (\citealt{Mason06,Mason09,Nikutta09,Ramos09,Ramos-Almeida2011a,Ramos-Almeida2011b,
Ramos14,Honig10,Herrero11,Herrero12a,Herrero13,Lira13}). Here we used the \citet{Nenkova08a,Nenkova08b} clumpy torus models, 
commonly known as CLUMPY, and the Bayesian tool
{\textit{BayesClumpy}} \citep{Asensio_and_Ramos09,Asensio_and_Ramos13} to fit the nuclear IR emission of NCG~2992.
The CLUMPY models are defined by six parameters (see Table \ref{tabtorus}), in addition to the
foreground extinction and the vertical shift required to match the model to the observed SED. A detailed description of the Bayesian
inference applied to the CLUMPY models can be found in \citet{Asensio_and_Ramos09}.

We constructed the nuclear IR SED of NGC\,2992 using the GTC/CC spectrum, which is more recent and has better angular resolution 
than the Gemini/MICHELLE one, extracted as a point source and resampled to 50 points; 
the UKIRT NIR nuclear fluxes from \citet{Herrero01}; the MICHELLE MIR fluxes from \citet{Ramos09}; and the 30~$\mu$m flux from the 
Spitzer/IRS spectrum (see Fig.~\ref{sed_torus_ngc2992}). There is good agreement between the flux calibration of  the nuclear spectrum
and the nuclear 11.2~$\mu$m flux, as we only measured a 10\% mismatch between them. 
For consistency, we scaled the spectrum to the nuclear 11.2~$\mu$m flux and we estimated a 15\% total uncertainty for the GTC/CC spectrum 
by quadratically adding the errors in the flux calibration and point source extraction.
We used the NIR nuclear photometry as upper limits because of the lower angular resolution of the UKIRT data. We did not use 
the HST/NICMOS image available in the archive for obtaining a NIR nuclear flux because the galaxy nucleus is saturated. 
We finally considered the IRS 30~$\mu$m flux as an upper limit, due to the low angular resolution of Spitzer.

%\textcolor{red}{which gets worse at such large wavelengths.}

%The process implies a decomposition (nucleus + bulge) of the radial
%surface brightness profiles and then applied to the nuclear flux a factor to isolate the nonstellar contribution, more details can be found in \citet{Herrero01}. One of the problems of this
%measure, in the particularly NGC\,2992 case, is that the HST NICMOS images are saturated (these images were used to know the nonstellar contribution) and the result can be overestimated. In the other hand, as
%we know NGC\,2992 is a galaxy with a huge variability and the observations could be made in a rise moment of the emission. 

%The clumpy torus model necessary parameters are the width of clouds angular distribution ($\sigma$), the radial extent of the tours (Y), the number of
%clouds along equatorial ray (N$_{0}$), the index of the radial density profile (q), the inclination angle of the torus (i), the optical depth
%per single cloud ($\tau_{\nu}$) and the foreground extinction (A$_{V}$). Thus ...
We fitted the SED of NGC\,2992 considering reprocessed torus emission and foreground extinction, using the IR extinction curve of \citet{Chiar_Tielens06}.  We used the prior A$_{V}^{(for)}$=[2,5] mag in our fit,
taking into account the extinction value of A$_{V}\sim$4 mag reported by \citet{Chapman00} for the innermost region of the galaxy. The openning angle of the ionization cones is 130$^\circ$, as measured 
from the [O~III] image reported by \citet{Garcia01}. This would correspond to a torus width of 25$^\circ$ and therefore we used the prior $\sigma$=[15$^\circ$,35$^\circ$] in our fit. We also used the prior i=[45$^\circ$,65$^\circ$] 
for the inclination angle of the torus, based on the values reported for the orientation of the accretion disk by \citet{Gilli00} using X-ray data, and from modelling of the kinematics of the NLR presented in \citet{Muller11}. The result of the 
fitting process are the posterior distributions of the parameters, but we can also translate the results into a best-fitting model, described by the combination of parameters that maximizes 
the posterior (maximum-a-posteriori; MAP) and a median model, computed with the median value of each posterior (see Fig. \ref{sed_torus_ngc2992}). The MAP and median parameters of NGC\,2992 are shown in Table \ref{tabtorus}.

From the fit presented in Fig. \ref{sed_torus_ngc2992} we find that we require a foreground extinction, unrelated to the torus,
of A$_{V}^{(for)}$ $\sim$5 mag in the case of the MAP model. This is in good agreement with the value  derived from 
the fit of the silicate feature reported by \citet{Colling11} using the Gemini/MICHELLE nuclear spectrum of NGC\,2992 ($\tau_{9.7\mu m}$~$\sim$0.3).

\begin{figure}
\centering
\includegraphics[width=8.85cm]{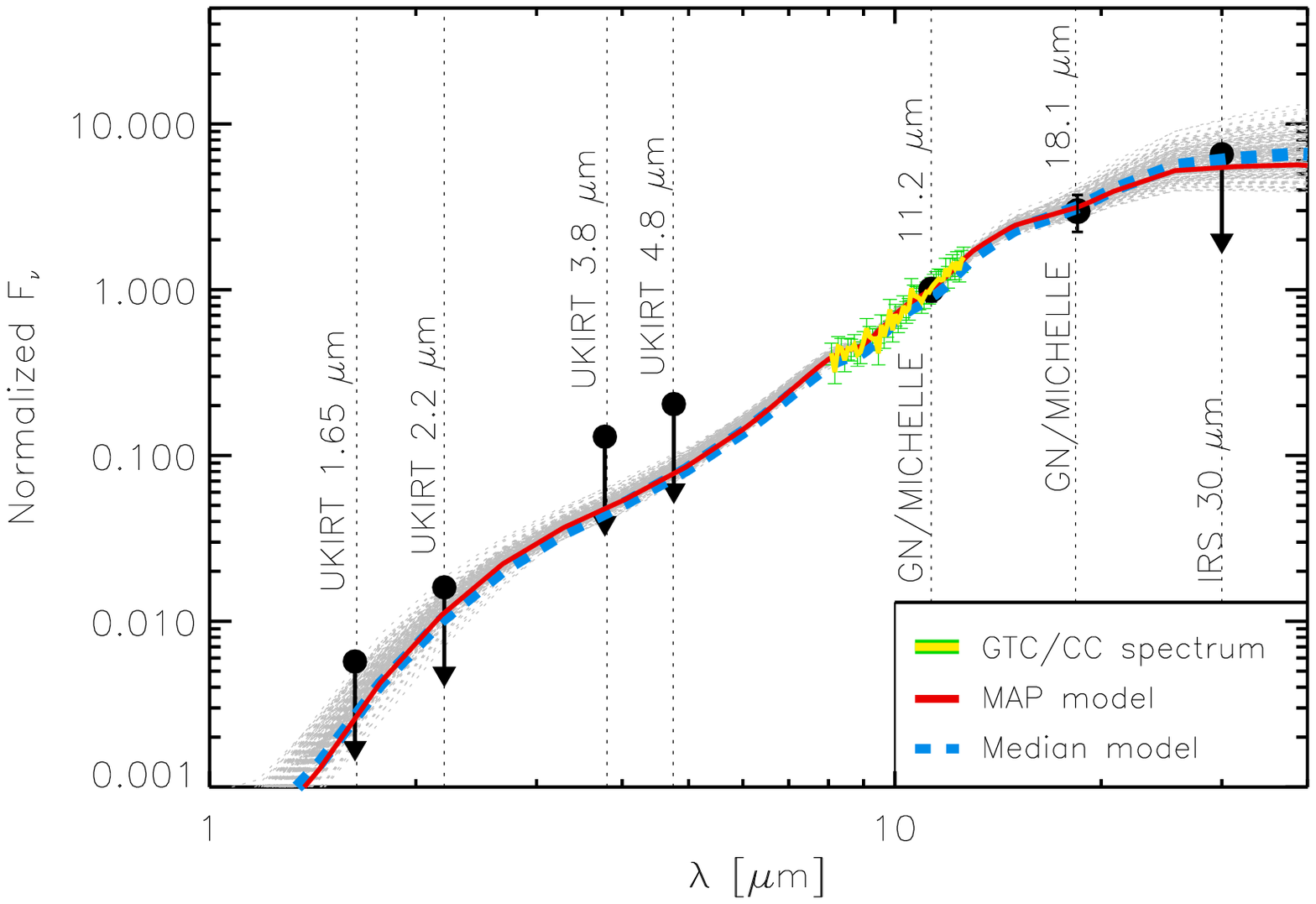}
\caption{Nuclear IR SED of NGC 2992 normalized at 11.2~$\mu$m (thick yellow line: GTC/CC nuclear spectrum; black dots: UKIRT NIR fluxes from \citet{Herrero01} and Gemini/MICHELLE 11.2, 18.1~$\mu$m fluxes from \citet{Ramos09} and 30~$\mu$m flux from the Spitzer/IRS spectrum). Solid red and dashed blue lines correspond to the MAP and median models respectively. Grey curves are the clumpy models sampled from the posterior and compatible with the data at 1$\sigma$ level.}
\label{sed_torus_ngc2992}
\end{figure}
\begin{table}
\centering
\begin{tabular}{cllc}
%\hline
\hline
Parameter &  Prior & Median & MAP \\
\hline
$\sigma$	 	 & [15\degr, 35\degr]	   & 34\degr$\pm1$\degr	& 35\degr    	\\
$Y$			 & [5, 100]		   & 16$\pm^{3}_{4}$  	& 12		\\
$N_0$			 & [1, 15]		   & 14$\pm1$  		& 15		\\
$q$		 	 & [0, 3]		   & 0.6$\pm^{0.3}_{0.5}$	& 0.5		\\
$i$		 	 & [45\degr, 65\degr]	   & 64\degr$\pm1$\degr	& 65\degr    	\\
$\tau_{V}$		 & [5, 150]		   & 105$\pm^{12}_{15}$  	& 109		\\ 
$A_V^{(for)}$			 & [2, 5] mag		   & 4.1$\pm^{0.9}_{0.5}$ mag	& 5.0 mag  	\\
\hline
R$_o$ 			& \dots          & 1.4$\pm^{0.5}_{0.4}$ pc                    & 1.2 pc          \\          
L$_{bol}^{AGN}/10^{43}$ & \dots          & 5.9$\pm^{1.2}_{0.9}~erg~s^{-1}$ & 5.8 $~erg~s^{-1}$   \\
$M_{\rm torus}/10^{5}$  & \dots          & 1.3$\pm^{0.3}_{0.2}\,M_\odot$   & 0.9$\,M_\odot$     \\
N$_H/10^{24}$			&\dots		&1.7$\pm{0.2}~cm^{-2}$&3.4$~cm^{-2}$\\
\hline      
\end{tabular}						 
\caption{Clumpy model parameters, intervals considered as uniform priors, median and MAP values of the posteriors resulting from the fit of NGC~2992 nuclear SED.
($\sigma$: width of clouds angular distribution; $Y$: radial extent of the torus; $N_0$: clouds along equatorial ray; $q$: index of the radial density profile;
$i$: inclination angle of the torus (Note: i=0\degr~is face-on and i=90\degr~is edge-on); $\tau_{V}$: optical depth per single cloud; $A_V^{(for)}$: foreground extinction). The last 4 rows 
correspond to the torus outer radius, the AGN bolometric luminosity, the torus gas mass and the hydrogen column density derived from the fit.}
\label{tabtorus}
\end{table}

We derived a small torus radius from the MAP torus model, of 1.2 pc, in agreement with the results from interferometry of nearby Seyfert galaxies (see \citealt{Burtscher13} and references therein). The torus covering factor is $\sim$0.5, which is more similar to the values reported by \citet{Ramos-Almeida2011a} for Seyfert 1 galaxies, and which could explain the variations in Seyfert type of NGC\,2992 (See Section \ref{disc_2992} for a 
discussion). We derived the AGN bolometric luminosity from the vertical shift applied to the MAP model to fit the data (see \citealt{Ramos-Almeida2011a} 
and \citealt{Herrero11} for further details), and we obtained L$_{bol}^{AGN}$= 5.8~x~10$^{43}$~erg~s$^{-1}$. This is consistent  
with the most recent X-ray observations available in the literature (L$_{bol}^{X-ray}$=3.2~x~10$^{43}$~erg~s$^{-1}$; see Table \ref{x_ray_fluxes}), taken in 2005. 
Using the MAP value of the total optical extinction produced by the torus (A$_V^{torus}$=1776 mag) we can derive the column density  using the dust-to-gas
ratio N$_H^{LOS}$=1.9~x~10$^{21}$ x A$_V^{torus}$ \citep{Bohlin78}. This gives N$_H$=3.4~x~10$^{24}$~cm$^{-2}$, which is within the range reported
by \citet{Weaver96} for the cold dense gas detected in the inner $\sim$3 pc of the galaxy in the X-rays ($\sim$10$^{23}-10^{25}$~cm$^{-2}$).
We also estimated the torus gas mass using equation 4 in \citet{Nenkova08b}, which is a function of the parameters $\sigma$, N$_0$, $\tau_V$, Y and the sublimation radius of the torus. We obtain a torus gas mass of M$_{torus}$=9~x~10$^{4}$ M$_\odot$, which 
is of the same order as that measured by \citet{Garcia14} for the central 20 pc of the Seyfert 2 NGC~1068 using cycle 0 data from the Atacama Large Millimeter/submillimeter Array (ALMA; M$_{gas}$=1.2~x~10$^{5}$ M$_\odot$).
%The 2-10 keV X-ray
%luminosity has changed over the years, because NGC\,2992 is a variable galaxy in the X-rays. 
The good agreement between L$_{bol}^{X-ray}$ and L$_{bol}^{AGN}$, in addition to the good match between the models and the IR observations, confirm that the clumpy torus fitted here provides a realistic scenario for the inner parsecs of NGC\,2992.

%The bolometric luminosity is quite similar L$_{bol}^{X-ray}$/L$_{bol}^{AGN}$=0.9, therefore we 
%concluded that the best fit is consistent with the expected bolometric luminositiy. Besides the bolometric luminosity, we derived a small torus radio from the MAP torus model of 1.74 pc.
\begin{table}
\centering
\begin{tabular}{lcll}
\hline
$L_{bol}^{X-ray}$ & Date & Telescope&Reference\\
($erg~s^{-1}$)& & &\\

\hline

2.3$\times10^{44}$& 1978 & HEAO~1& \citet{Mushotzky82}\\
8.9$\times10^{43}$& 1979 & Einstein& \citet{Turner91}\\
2.5$\times10^{43}$& 1990 & Ginga& \citet{Nandra1994}\\
1.4$\times10^{43}$& 1994 & ASCA& \citet{Weaver96}\\
1.2$\times10^{43}$& 1997 & BeppoSAX~1& \citet{Gilli00}\\
1.4$\times10^{44}$& 1998 & BeppoSAX~2& \citet{Gilli00}\\
2.6$\times10^{44}$& 2003 & XMM-Newton& \citet{Shu10}\\
2.5$\times10^{44}$& 2003 & XMM-Newton& \citet{Brightman11}\\
3.2$\times10^{43}$& 2005 & Suzaku& \citet{Yaqoob07}\\
\hline
\end{tabular}						 
\caption{NGC 2992 bolometric luminosities obtained from the 2-10 keV X-ray intrinsic luminosities (by multiplying by a factor 20 as in \citealt{Elvis94}). For the non-intrinsic X-ray luminosity values, a factor 70 has been used in addition to the
bolometric correction factor, as in \citet{marinucci12}.} 
%\textcolor{blue}{This correction factor is used for Compton-thick sources (N$_{H}>$10$^{24}$~cm$^{-2}$) for more details see \citealt{marinucci12} and references therein.}
\label{x_ray_fluxes}
\end{table}

%with it is consistent with the spectral decomposition. Since, we can made a good reproduction of the MIR
%Spitzer/IRS spectrum, with the AGN here derived and SB template

\subsection{Large scale emission}
\label{large}
In this section we study the large scale IR morphology of the interacting system Arp\,245 (see Fig. \ref{ngc2992__all}), 
which consists of three galaxies undergoing strong tidal interaction: the spirals
NGC\,2992 and NGC\,2993 and the tidal dwarf galaxy Arp\,245N. There is a fourth galaxy that is also part of the system: 
FGC\,0938, but it lies outside the Spitzer and Herschel FOVs.

\begin{figure*}
\centering
\includegraphics[width=14.cm, angle=90]{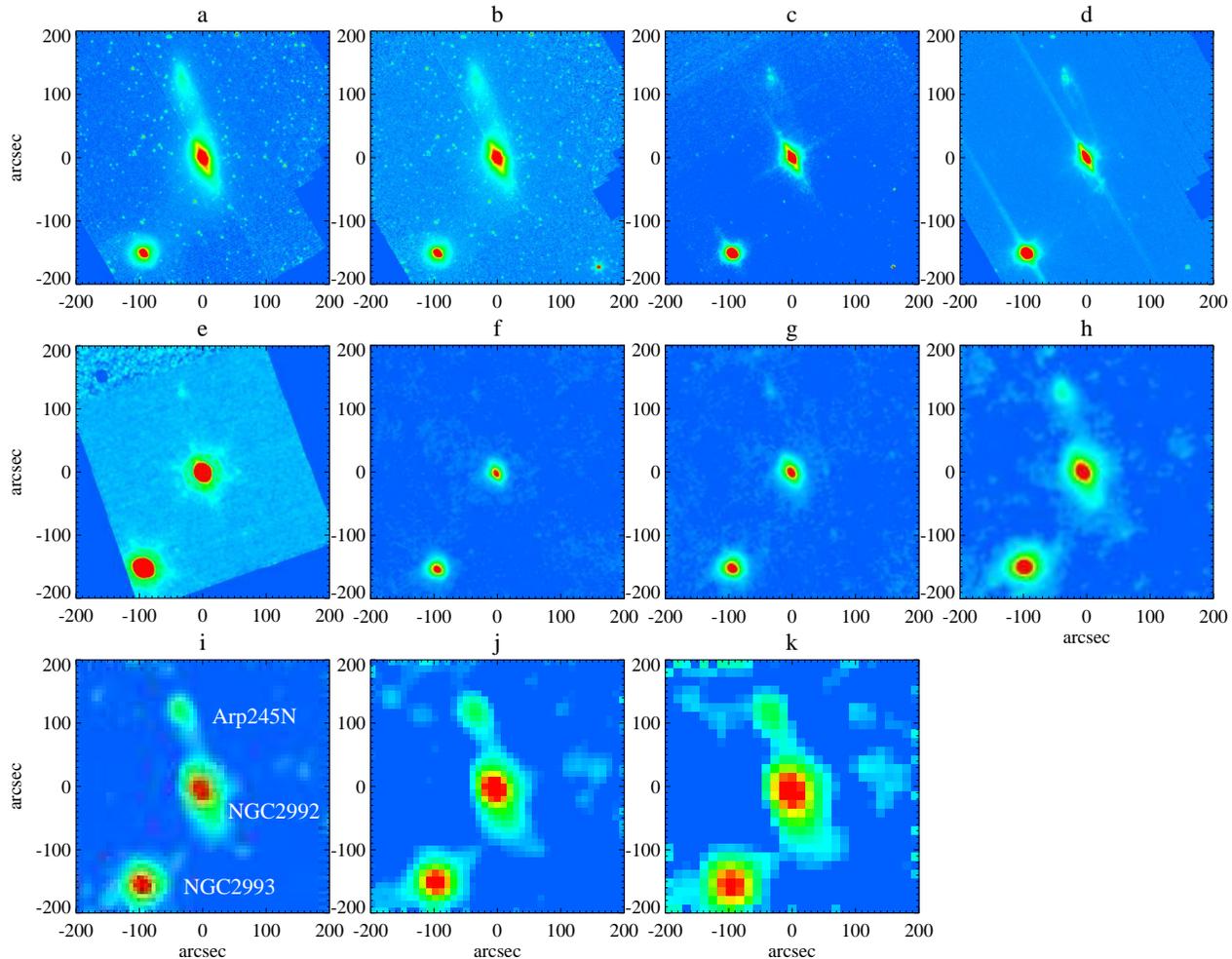}
\caption{Spitzer and Herschel images of the Arp 245 system. NGC\,2992 is the central galaxy, NGC\,2993 is at the bottom 
left corner of the images, and Arp\,245N is at the top.
The four IRAC channels  (3.6, 4.5, 5.8 and 8~$\mu$m) are shown in panels a-d, the MIPS channel 1 image (24~$\mu$m) 
in panel e, the PACS 70, 100 and 160~$\mu$m images in panels f-h, and the SPIRE 250, 350 and 500~$\mu$m images in panels i-k. 
All images have been smoothed (3 pixels box). North is up, and east to left.}
%NGC\,2992(center), NGC\,2992 North (top) and NGC2993(bottom). The Fig. shown the different wavelength images. The different channels of IRAC instrument (a) 
%3.6 micron, (b) 4.5 micron, (c) 5.8 micron and (d) 8 micron, the MIPS channel 1 image (d) 24 micron, the different filters of PACS (e) 70  micron, (f) 100 micron and 
%(g) 160 micron, and, finally, the filters of SPIRE (i) 250 micron, (j) 350 micron and (k) 500 micron. All images are smooth with a box of 3 pixels. {\bf{Note:}} North is up and east is left.}
\label{ngc2992__all}
\end{figure*}

\subsubsection{NGC 2992}
\label{large_2992}

In Fig.~\ref{ngc2992_spitzer_center} we show Spitzer/IRAC~\&~MIPS and Herschel/PACS~\&~SPIRE images of NGC\,2992 in a 
120\arcsec$\times$120\arcsec~FOV.  The extended emission in the IRAC images is elongated in the same direction
as the faint extended emission detected in the Gemini/MICHELLE N-band and HST/NICMOS images (see Fig. \ref{hst_vs_gemini}), 
and it also coincides with the orientation of the galaxy major axis. Considering that the spectrum of the extended 
emission shows a strong 11.3 \micron~PAH feature (see Section \ref{spectroscopy}), the bulk of 
this faint emission is likely produced by dust heated by star formation. In the case of the IRAC images the source 
of the IR emission depends on the band we look at, 
with the 3.6 and 4.5 \micron~emission 
likely dominated by starlight and the 5.8 and 8 \micron~emission by dust heated by star formation and the AGN (e.g. \citealt{Draine07}).

%,{\bf{ specially in the 
%5.8 and 8~$\mu$m images. Considering that these wavelengths trace interstellar dust and SF, the}} 
%This indicates that the origin of this faint extended component 
%This indicates that the 
%bulk of this faint extended emission is likely dust in the inner galaxy disk with some contribution
%from SF (see Section \ref{spectroscopy}), as its orientation coincides with the major
%axis of the projection of the galaxy in the sky. %Besides, the biconical ionization
%structure reported by \citet{Allen99} is aligned perpendiculary to the faint IR emission, ruling out a possible contribution from the NLR.

%The 3.6 and 4.5~$\mu$m~IRAC images are good extinction-free tracers of the stellar light \citep{Regan04}.
%Whereas, the 5.8~$\mu$m~IRAC image covers the wavelength range that includes the 6.2~$\mu$m PAH line and the 8~$\mu$m channel includes the 7.7 and
%8.6~$\mu$m PAH emission features \citep{Howell07}. Therefore, both channels are tracing SF activity and interstellar dust emission. Stellar continuum emission 
%is insignificant compared to PAH emission in the IRAC 8~$\mu$m band in nearly all nearby 
%LIRGS \citep{Mazzarella07}. Mazzarella07,submitted? }}

The MIPS 24~$\mu$m image, on the other hand, is rather point-like, as the Gemini/MICHELLE Q-band emission seen in the 
right panel of Fig. \ref{Np_smooth}.
The 70 and 100~$\mu$m emission seen in the Herschel/PACS images (panels f and g in Fig. \ref{ngc2992_spitzer_center}) 
is barely resolved and slightly elongated in the 
same direction as the Spitzer/IRAC images. From 160 to 500~$\mu$m, the intensity of the extended emission increases 
significantly, produced by cooler dust in the galaxy 
(panels h-k in Fig. \ref{ngc2992_spitzer_center}).

\begin{figure}
\centering
\includegraphics[width=9.cm]{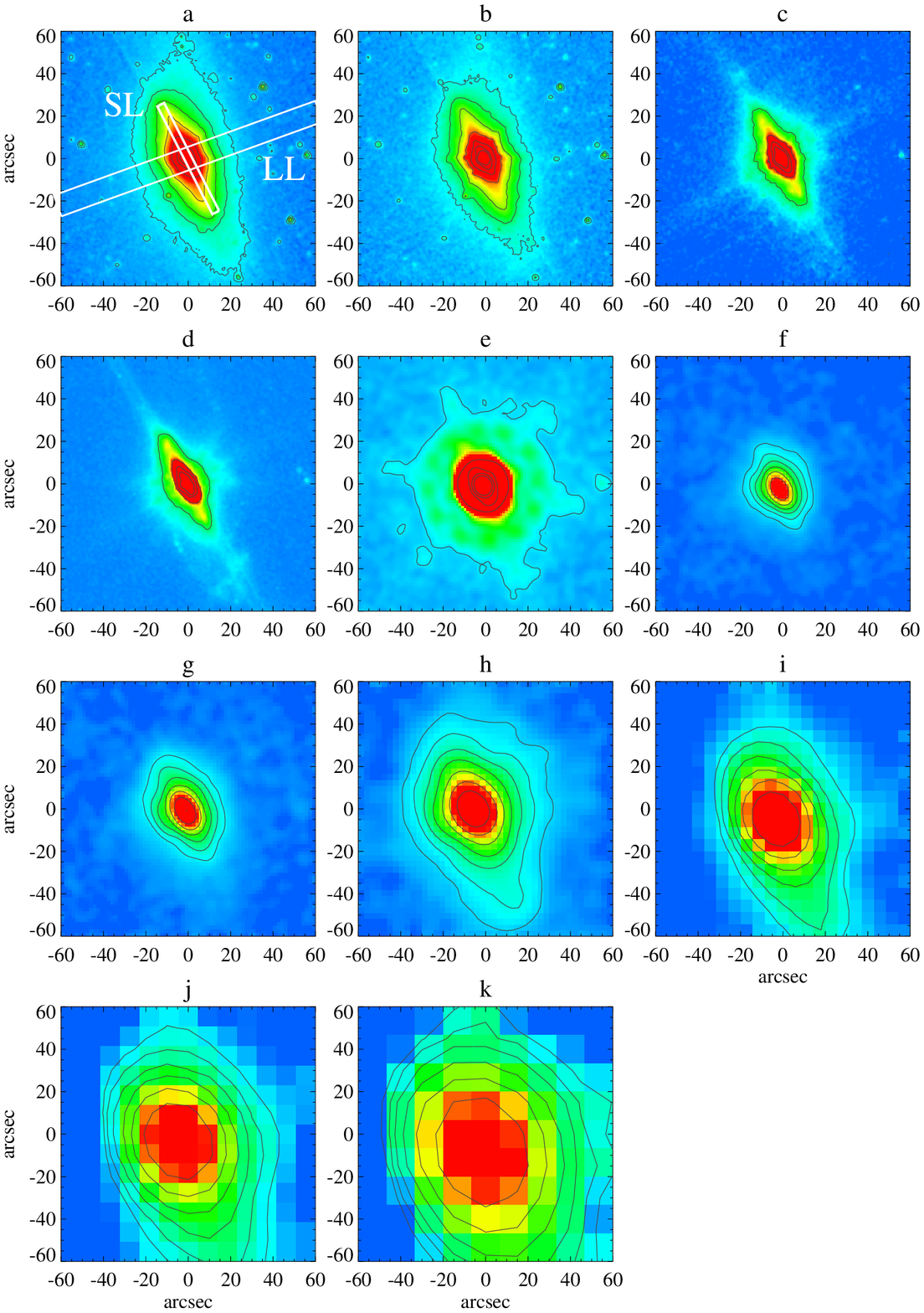}
\caption{Same as in Fig. \ref{ngc2992__all}, but for NGC\,2992. Panel (a) shows the orientation of the Spitzer/IRS slits. All images have been smoothed (3 pixels box) and have their own contours overlaid (in black).}
\label{ngc2992_spitzer_center}
\end{figure}

\subsubsection{Arp 245 system}
\label{large_245}

In Fig. \ref{ngc2992__all} we show Spitzer/IRAC \& MIPS and Herschel/PACS \& SPIRE images of the interacting system Arp\,245. 
NGC\,2992 and Arp\,245N show 
clearly distorted morphologies, and they are linked by a tidal tail which appears very bright in the Spitzer/IRAC images. The
bridge (Arp\,245 Bridge) between NGC\,2992 and NGC\,2993 is more conspicuous in the Herschel/SPIRE maps, indicating the presence 
of cooler dust.

\begin{table*}
\scriptsize
%\tiny
\centering
\begin{tabular}{lrrrrrrrrrrr}

\hline
 &3.6$\mu$m&4.5$\mu$m&5.8$\mu$m&8$\mu$m&24$\mu$m&70$\mu$m&100$\mu$m&160$\mu$m&250$\mu$m&350$\mu$m&500$\mu$m \\
\hline

NGC~2992& 180$\pm$19& 158$\pm$19& 244$\pm$25 & 510$\pm$53& 1000$\pm$108& 9070$\pm$1014& 12018$\pm$1344& 10794$\pm$1207& 3882$\pm$443& 2104$\pm$240& 357$\pm$41\\

NGC~2993& 55$\pm$6& 41$\pm$4& 132$\pm$14& 527$\pm$60& 1182$\pm$127& 12138$\pm$1357& 13880$\pm$1552& 10776$\pm$1205& 3291$\pm$376& 1714$\pm$196& 291$\pm$33\\

Arp~245N&  7$\pm$1 &4$\pm$1&8$\pm$1&22$\pm$2&3$\pm$1& 77$\pm$9&163$\pm$18&394$\pm$44 &279$\pm$32&176$\pm$20&31$\pm$4\\
%NGC 2992& 1 FWHM&3.6& 130\\
%NGC 2992& 1 FWHM&4.5& 123\\
%NGC 2992& 1 FWHM&5.8& 191 \\
%NGC 2992& 1 FWHM&8& 371\\
%NGC 2992& 1 FWHM&24& 868\\
%NGC 2992& 1 FWHM&70& 7322\\
%NGC 2992& 1 FWHM&100& 9364\\
%NGC 2992& 1 FWHM&160& 7397\\
%NGC 2992& 1 FWHM&250& 2251\\
%NGC 2992& 1 FWHM&350& 918\\
%NGC 2992& 1 FWHM&500& 96\\
\hline
\end{tabular}						 
\caption{Total fluxes measured for NGC~2992, NGC~2993 and Arp~245N in mJy. The flux errors are the result of adding quadratically
the calibration uncertainties [3\% for IRAC \citep{IRAC13}, 4\% for MIPS at 24~$\mu$m \citep{MIPS11}, 5\% for PACS \citep{PACS13} 
and 5.5\% for SPIRE \citep{SPIRE14}] and the estimated uncertainty for aperture photometry ($\sim$10\%). Note that in the case of SPIRE,
extended sources have an additional 4\% error due to uncertainty in the measurement of the beam area.}
\label{flux_results}
\end{table*}

We measured total galaxy fluxes for the three galaxies to construct their IR SEDs, which are shown in Fig. \ref{SED} and 
Table \ref{flux_results}. 
We used a large aperture (60\arcsec~radius for NGC\,2992/93 and 25\arcsec~radius for Arp\,245N) to collect all 
the galaxy flux\footnote{In the case of the MIPS 24 \micron~image, we used an aperture radius of 25\arcsec~to calculate
the total flux of NGC\,2993, which is too close to the image edge.}, and subtracted the sky background.
%{\bf We used
%a large aperture, of 60\arcsec~radius, to collect all the galaxy flux\footnote{In the case of the MIPS 24 \micron~image, 
%we used an aperture radius
%of 25\arcsec~to calculate the total flux of NGC\,2993, which 
%is too close to the image edge.}, and sky subtraction was perfomed. In the case of Arp\,245N, we used a smaller aperture radius of 
%25\arcsec, enough to include all the galaxy emission.} 
The IR SEDs of NGC\,2992/93 shown in Fig. \ref{SED} also include NIR total fluxes from 2MASS, reported by \citet{Jarrett03}.

\begin{figure}
\centering
\includegraphics[width=6.2cm,angle=90]{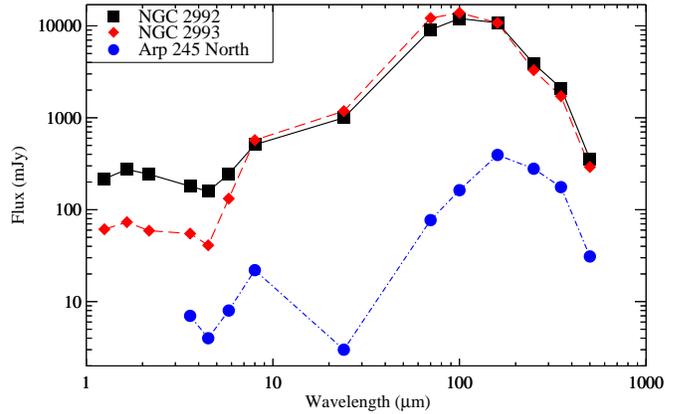}
\caption{NGC\,2992, NGC\,2993 and Arp\,245N total IR SEDs (black solid, red dashed and blue dot-dashed lines respectively).}
\label{SED}
\end{figure}

As can be seen from Fig. \ref{SED}, the total SEDs of NGC\,2992 and NGC\,2993 are very similar at wavelengths 
longer than 8 \micron.
%, in spite of the fact that NGC\,2992
%is a Seyfert galaxy and NGC\,2993 a {\bf{non-active}} star forming galaxy. The similarity is likely related to the presence of intense
%SF in NGC\,2993, which heats the dust at the same temperature as the AGN does in NGC\,2992. This
%is consistent with the scenario reported by \citet{duc00}: they simulated numerically
%the interaction undergone by the galaxy system and concluded that the past galaxy system collision distributed a similar amount of 
%dust to both spiral galaxies and also turned on the nuclear activity of NGC\,2992 and a SB in NGC\,2993 at the same
%time \citep{duc00}. 
This similarity is likely related to the presence of intense SF in both galaxies,
which heats the dust at similar temperatures. This is corroborated by 
the SED fitting presented in Section~\ref{dust}. On the other hand, the total SEDs are different 
in flux level at shorter wavelengths ($\lambda<$8~$\mu$m), 
with NGC\,2992 being brighter.
Considering that the 3.6 and 4.5 \micron~emission is dominated by the Rayleigh-Jeans tail of stellar photospheric 
emission \citep{Howell07}, the brighter NIR SED of NGC\,2992 is likely due to its larger stellar mass, 
as compared to that of NGC\,2993 \citep{duc00}, and, to a lesser extent, to the extra-contribution
from dust heated by the AGN. Nevertheless, we note that the SED shapes at these shorter wavelengths 
are very similar, and characteristic of old stellar populations.

%(M$_{H_2}$=1.19~x~10$^9$~M$_{\odot}$
%and 0.39~x~10$^9$~M$_{\odot}$ respectively; \citealt{duc00}). %\textcolor{blue}{This difference in the H$_2$ content is due to the existence of a SB in the NGC\,2993 galaxy,
%which consumes a lot of molecular gas, whereas in NGC\,2992 Seyfert galaxy there is an active nuclei that caused the inexistence of a SB.}
%is more massive than 
%NGC\,2993~(M$_{H I}$=1.05$\cdotp$10$^9$~M$_{\odot}$ and M$_{H_2}$=0.39$\cdotp$10$^9$~M$_{\odot}$), which values have been reported by \citet{duc00}.
%This is probably due to the AGN contribution in addition to the stellar emission. 
%Both spiral galaxies appear unresolved in the Spitzer/MIPS image, and we do not detect any extended emission. 

%We find the lowest IR flux at 4.5~$\mu$m for both galaxies, which is due to the CO absorption band at $\sim$4.7~$\mu$m \citep{Lutz04}. %This can be due to the poorer sensitivity of the MIPS instrument, as compared to IRAC.

%{\bf{\citep{Mitchell89}}}.

The fainter SED of Arp\,245N has a similar shape to those of NGC\,2992 and NGC\,2993 SEDs, except in the $\sim$24-100~$\mu$m range, 
where the fluxes are lower compared to other wavelengths. This can be indicative
of a relatively weak illuminating radiation field that may possibly be consistent with weak SF activity (see Section \ref{dust}).
We refer the reader to \citet{DraineandLi07} and \citet{Dale2014} for further discussion.

\section{Recovering nuclear information from arcsecond resolution data}
\label{recover}

In this section we used different methods to try to recover the nuclear emission of NGC\,2992 from the lower angular resolution 
data of Spitzer and Herschel. This kind of analysis is important due to the paucity of ground-based MIR instruments and to the difficulty 
of observing in the MIR range from ground. In this case, we have the opportunity to compare with the nuclear fluxes obtained 
from the high angular resolution data presented here, obtained using GTC/CC and Gemini/MICHELLE.

%O. In this section we used different methods  
%to try to recover the information provided by high angular resolution data used here (GTC/CC and GN/MICHELLE) from low as those angular resolution data from Spitzer and Herschel. 

%emission is dominant (20-25~$\mu$m) the image is quite punctual.

\subsection{Recovering the nuclear IR SED}
\label{nuclear_big}

%In this section we use two different methods to try to recover the nuclear IR SED of NGC\,2992 from the lower angular resolution from Spitzer and Herschel data.\\

In order to try to recover the nuclear IR SED of NGC\,2992 from low angular resolution data, we used two different methods, which are
described below. 

\subsubsection{Aperture Photometry}
\label{aperture_phot}

We used aperture photometry of the galaxy nucleus in the Spitzer and Herschel images. The photometry 
was carried out with the {\textit{DIGIPHOT}} package of {\textit{IRAF}}. We used
different apertures (multiples of the FWHM in each band;~see Table \ref{fluxes}) and 
%and performed sky subtraction, which was carried out using a concentric ring large enough to contain a good estimate of the sky background and we applied corresponding aperture photometry corrections. To calculate 
%the aperture corrections, we used 
the sky subtraction was carried out using a concentric ring large enough to exclude the galaxy emission. We finally applied corresponding aperture corrections in order to recover the unresolved galaxy flux. These correction factors were computed from
the extended Point Response Functions (PRFs) for IRAC\footnote{http://irsa.ipac.caltech.edu/data/SPITZER/docs/}, the core PRF for MIPS$^2$ and synthetic PSFs for Herschel/ PACS\footnote{ftp://ftp.sciops.esa.int/pub/hsc-calibration/PACS/PSF/}
%{https://nhscsci.ipac.caltech.edu/sc/index.php/Pacs/PSFs}
 and SPIRE\footnote{ftp://ftp.sciops.esa.int/pub/hsc-calibration/SPIRE/PHOT/}. We performed aperture photometry on 
the PRFs and PSFs stars, using the same apertures as for the galaxy in each band, and calculated the different correction 
factors by comparing each individual value with the total flux, measured in an aperture large enough to contain all the star 
flux. Note that the synthetic PSFs have different pixel scales than the science data and we used the {\textit{MAGNIFY}} task of {\textit{IRAF}} 
to resample the PSF images.

\begin{table*}
\centering
\begin{tabular}{lccccccccc}
\hline
Instrument/& Wavelength &FWHM & FWHM &Aperture& Flux& 100\%~PSF& \multicolumn{2}{c}{Best PSF} & Flux\\
Band&$\lambda_{c}$ & & &photometry&uncertainty&substraction  & \multicolumn{2}{c}{substraction}&uncertainty\\
& ($\mu$m) &(arcsec)& (pc)& flux & (percent)& flux & (percent) & flux& (percent) \\
\hline
IRAC/Ch1&3.6&	1.85& 322	&61&10& 59	& 95	&   56& 15 \\
IRAC/Ch2&4.5 &	1.77& 308	&78&10& 76	& 95	&   73& 15 \\
IRAC/Ch3&5.8&	2.15& 374	&148&10& 148	& 80	&  119& 15 \\
IRAC/Ch4&8.0&	2.79& 485	&261&10& 272	& 50	&  136& 15 \\
MIPS/Ch1&24&	6.13& 1067	&874&11& 905	& 90	&  815& 16 \\
PACS/Blue&70&	5.25& 914	&5852&11 & 3929	& 60	& 2357& 16 \\
PACS/Green&100&	6.75& 1175	&8080&11& 5417	& 60	& 3250& 16 \\
PACS/Red&160&	10.80&1879 	&7910&11& 6135	& 80	& 4908& 16 \\
SPIRE/PSW&250&	17.63& 3068	&2805&11& 2446	& 65	& 1590& 16 \\
SPIRE/PMW&350&	24.49&	4261	&1608&11& 1607	& 70	& 1125& 16 \\
SPIRE/PLW&500&	34.66&	6031 	&287&11&  263	& 75	&  197& 16 \\
\hline
\end{tabular}						 
\caption{NGC~2992 nuclear fluxes calculated using aperture photometry and PSF subtraction (all fluxes in mJy units). Column 1 and 2 list the instrument/band and its central wavelenght ($\mu$m). 
Column 3 and 4 correspond to each band resolution in arcsec and parsecs. Columns 5 and 6 list the aperture photometry fluxes and corresponding errors, with the
aperture photometry carried out in an aperture equals to 2 FWHM (See Section \ref{aperture_phot}). Column 7 corresponds to the fluxes obtained from 100\% PSF substraction, and columns 8, 9 and 10 list the
best percentage of PSF subtraction used in each band , corresponding fluxes and errors. 
The flux errors are the result of adding quadratically the flux calibration uncertainties (see Table \ref{flux_results}) and the errors associated with either aperture photometry ($\sim$10\%) 
or PSF subtraction ($\sim$15\%).}
\label{fluxes}
\end{table*}

We took multiples of the FWHM and then we applied aperture corrections for each one. We find that the aperture corrections 
converged around 2 FWHM, and then chose this aperture to estimate the nuclear fluxes.
%due to one FWHM proved to be too small (few pixels) to provide correct aperture corrections and thus
%to provide correct nuclear fluxes. Therefore, we finally 
%chose an aperture of 2 FWHM to estimate the nuclear fluxes.
In Table~\ref{fluxes} and Fig.~\ref{nuclear_sed} we show the nuclear galaxy fluxes calculated 
using this method and including aperture corrections.

\begin{figure}
\centering
\includegraphics[width=6.cm,angle=90]{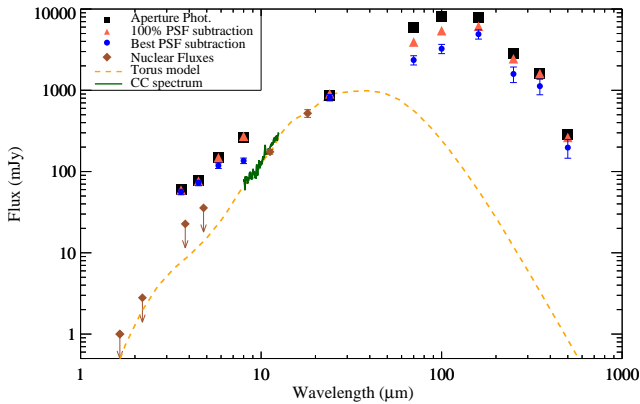}
\caption{Nuclear fluxes of NGC\,2992 from low angular resolution data calculated using aperture photometry (squares) and PSF subtraction at various levels (triangles and circles). The nuclear IR SED (diamonds), including the GTC/CC spectrum and the torus model fitted in Section \ref{bayesclumpy}, are shown for comparison.}
\label{nuclear_sed}
\end{figure}

\subsubsection{Subtraction of scaled PSFs}
\label{psf_sub}

As the aperture photometry might include contamination from the foreground galaxy, here we use the PRF and PSF stars to 
obtain more realistic nuclear fluxes. First, we scaled the PSF stars to the peak of the galaxy emission in each band, which represents 
the maximum contribution of the unresolved source, and we integrated 
the flux in an aperture large enough to contain all the star flux. Then, the host galaxy contribution corresponds to the
total galaxy emission minus the scaled PSF (i.e. the residual of the subtraction). We require a relatively flat profile
in the residual for a realistic galaxy profile and therefore reduce the scale of the PSF from matching the peak
of the galaxy emission to obtain the unresolved fluxes, as in \citet{Radomski2002} and \citet{Ramos09}. Fig. \ref{nuclear_psf_sub} 
shows an example of PSF subtraction at various levels (in 3$\sigma$ contours) 
for the 160~$\mu$m Heschel/PACS image. In this case 80\% PSF subtraction produces a flatter profile than 90 and 100\%, 
which are clearly over-subtracted. 

The nuclear fluxes calculated using this method are shown in Table \ref{fluxes} and 
Fig. \ref{nuclear_sed}. We also included the results from 100\% PSF subtraction, which are more similar to the aperture 
photometry fluxes, calculated in Section \ref{aperture_phot}. 
We note that the relatively flat profile required to determine the best PSF subtraction level might not be completely 
flat due to possible diffuse or irregular dust structures \citep{munoz-mateos09}. This potential issue could 
be affecting the fluxes calculated using this method.

\begin{figure}
\centering
\includegraphics[width=7.0cm]{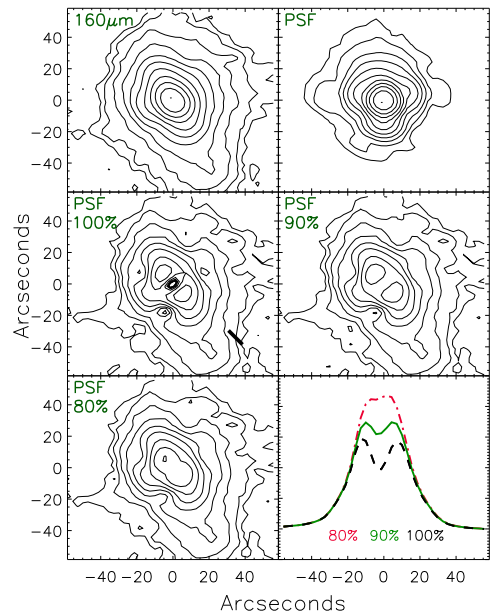}
\caption{160~$\mu$m Heschel/PACS contours of NGC\,2992, the PSF star and the scaled PSF subtraction at the 80, 90 and 100\% levels 
(red dot-dashed, solid green and black dashed lines respectively). The best subtraction is 80\%, according to the flat galaxy 
profile shown in the bottom-right panel. North is up, and east to left.}
\label{nuclear_psf_sub}
\end{figure}

In Fig. \ref{nuclear_sed} we compare the high angular resolution SED and the MAP torus model (see Fig. \ref{sed_torus_ngc2992}) 
with those from aperture photometry and PSF subtraction. The latter includes 100\% and best PSF subtraction, where the percentage 
of the best PSF subtraction corresponds to the one that produces the flattest galaxy profile in each band. The best PSF subtraction 
fluxes are the smallest, but still significantly larger than the high angular resolution SED. Therefore, we find that we cannot 
recover the nuclear SED from the Spitzer and Herschel data, with the exception of the 24~$\mu$m flux, which we recover with both 
methods. This is expected, since at 20-25~$\mu$m the torus emission is supposed to dominate (see \citealt{Ramos-Almeida2011a} 
and references therein). At shorter and longer wavelengths the galaxy contribution increases and contaminates our aperture 
photometry and PSF subtracted fluxes.

%\subsubsection{Method: GALFIT decom\bibitem[Alonso-Herrero et al.(2013)]{Herrero13} Alonso-Herrero, A., et al. 2013, ApJ, 779, L14position}
%\label{galfit_photometry}
%We wil try to get more realistic nuclear fluxes.....

\subsection{Spectral Decomposition}
\label{decomposition}

\begin{figure}
\centering
\par{
\includegraphics[width=6.5cm,angle=90]{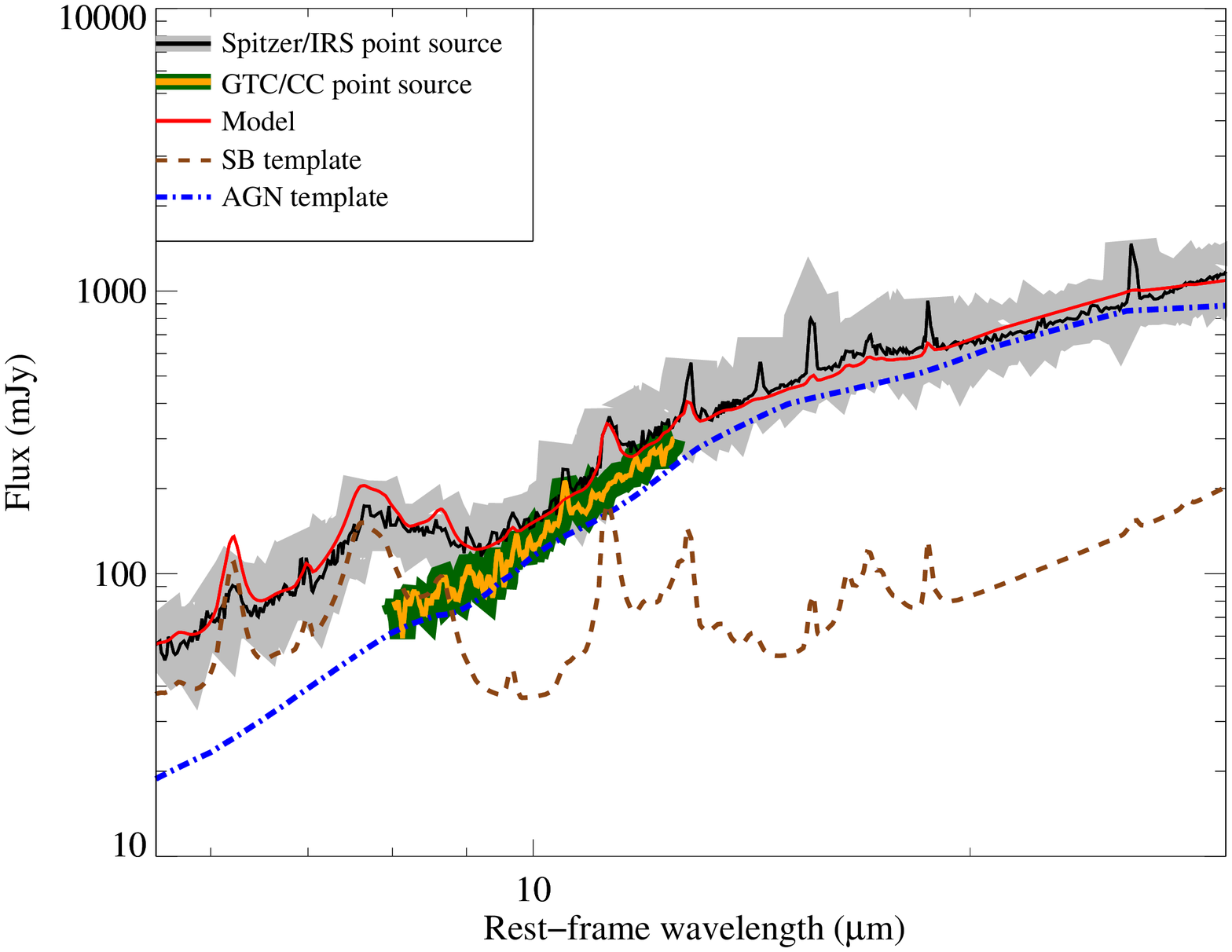}
\includegraphics[width=6.5cm,angle=90]{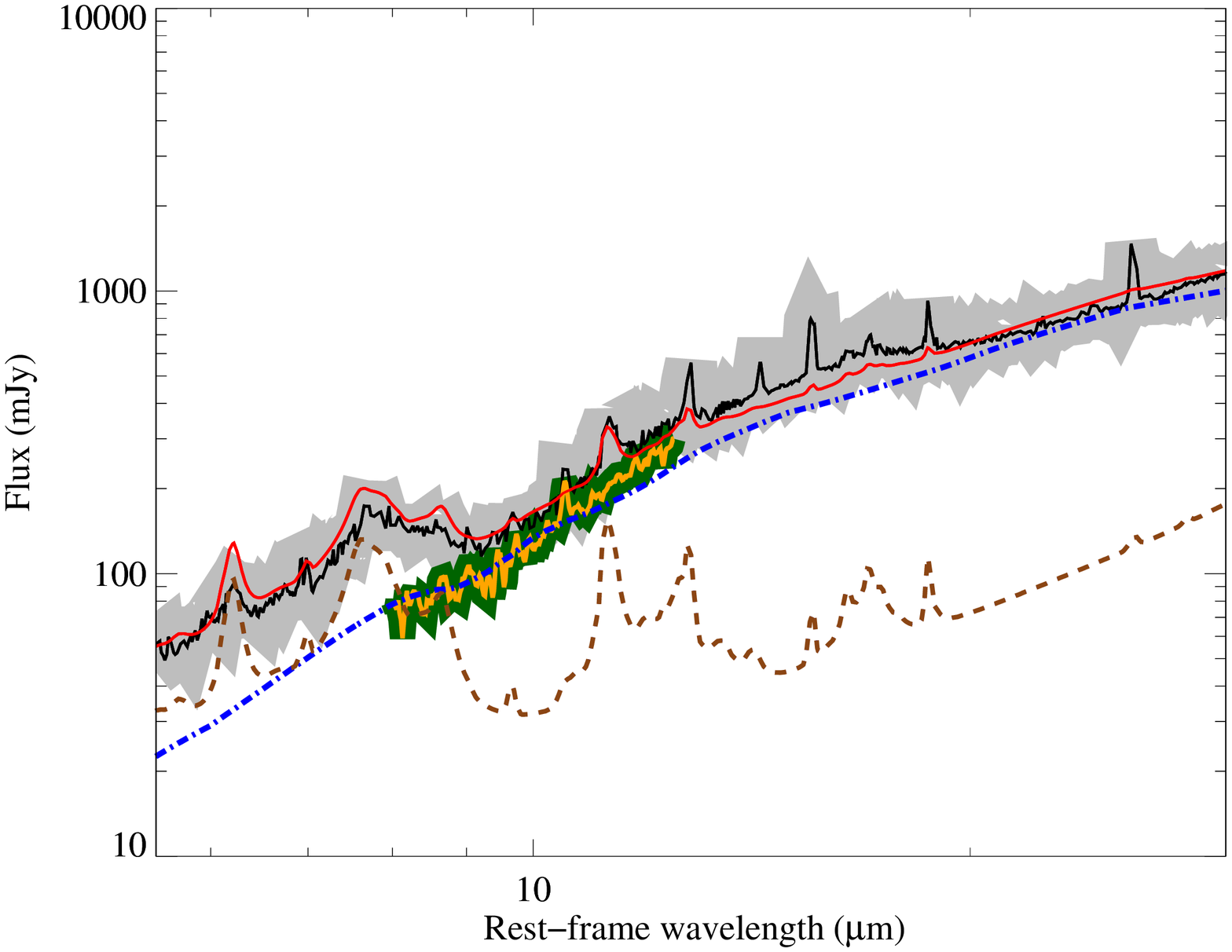}\par} 
\caption{NGC\,2992 6-30~$\mu$m Spitzer/IRS rest-frame spectrum (black solid line), best fit (red solid line), MAP torus model 
(blue dot-dashed line), and SB template (brown dashed line). Top panel: Fit using the MAP torus model fitted to the high angular 
resolution IR SED of NGC\,2992. Bottom panel: Fit using the average Seyfert 2 torus model fit from \citet{Ramos-Almeida2011a} and 
applying an iterative process. The 8-13~$\mu$m GTC/CC rest-frame spectrum is shown in both panels for comparison (smoothed with a 
2 pixel box; orange line). The uncertainties of the Spitzer/IRS and GTC/CC spectra are shown as grey and dark green 
shaded regions respectively.}
\label{resta_agn}
\end{figure}

Considering the spatial scales probed by the Spitzer/IRS spectrum of NGC\,2992 ($\sim$630 pc) and the prominent 11.3~$\mu$m PAH feature shown in Fig. \ref{spec_ngc2992}, we expect contributions from the AGN and SF on these scales. To estimate the AGN contribution to the Spitzer/IRS spectrum, we take the simple approach of decomposing it in AGN and SB components. To do that, we used the average spectrum of local SB of \citet{Brandl06} and the templates of purely star-forming LIRGs
of \citet{Rieke09}, which cover the IR luminosity range 10$\leq$log($L_{IR}/L_{\odot}$)$\leq$12. As AGN template we used the MAP clumpy torus model (i.e. the best-fitting model) fitted in Section \ref{bayesclumpy}, and shown in Fig. \ref{sed_torus_ngc2992}. The fitting procedure is described in detail in \citet{Herrero12b}.

%Taking advantage of the IR nuclear SED and of the nuclear 
%spectrum observed from GTC/CC ($\sim$90 pc), we estimated the AGN contribution to the MIR emission using the previous SED modelling %with clumpy torus models.

%In these type of decompositions are typical using the two average AGN templates \citep{Ramos-Almeida2011a} of the large
%database of the CLUMPY models and for the star-forming template, the average spectrum of local SB of \citet{Brandl06} and the templates %of purely star-forming (U)LIRGs
%of \citet{Rieke09} covering the IR luminosity range 10$\leq$log($L_{IR}/L_{\odot}$)$\leq$12. A similar procedure is describe in detail %in \citet{Herrero12b}. However, they used an iterative method 
%to perform the decomposition and to find the best AGN component. Nevertheless, we have privileged information from high angular %resolution data, therefore, we used the best AGN fitting
%(the previous clumpy torus model) in our decomposition (following the same procedure that \citet{Ramos14}).
The fit was carried out in the spectral range 6-30~$\mu$m to avoid the slightly decreased signal-to-noise 
of the longest wavelengths.
We tried different combinations of
AGN + SB templates, allowing rescaling of the two components. We finally chose the fit that minimized the $\chi^2$, which in this case was 
the SB template of IR luminosity log($L_{IR}/L_{\odot}$)$=$10 of \citet{Rieke09}, in combination with the AGN. This IR luminosity is indeed similar to that
of NGC\,2992 (log$(L_{IR}/L_{\odot})= 10.52$; \citealt{Sanders03}). The result of the fit is shown in the top panel of Fig. \ref{resta_agn}. We have quantified 
the AGN fractional contributions to the total 6, 20 and 30~$\mu$m emission and obtained 34, 88 and 81\% respectively. We note that 
the scaled AGN template concides with the nuclear GTC/CC spectrum within the errors, 
proving the reliability of the method employed here. %and the great advantage that implies the privileged information from high angular resolution spectroscopy. 

By using this method we are taking advantage of the privileged information of the high angular resolution GTC/CC spectrum. Therefore, we repeated the same
process without using this information and we tried to recover the AGN and SF contributions using the average Seyfert 2 CLUMPY 
torus model fit reported by \citet{Ramos-Almeida2011a} as initial AGN template. In this case the fitted AGN component does not
match the high angular resolution GTC/CC spectrum and therefore it is 
necessary to perform an iterative process to obtain a reliable fit. After the first fit, we subtracted the SB template from the Spitzer/IRS spectrum,
and then fitted the residual with BayesClumpy. We then performed a second fit using the latter MAP model and the SB templates and repeated the 
process until the solution converged. To quantify that, we used the AGN fractional contributions at the wavelengths listed before. The percentages
at 20 and 30~$\mu$m converged after the fourth iteration, but at 6~$\mu$m convergence is not reached. This is due to the prominent PAH features at
6.2, 7.7 and 8.6~$\mu$m of the SB template, which results in a overestimation of the PAHs contribution to the 
Spitzer/IRS spectrum at these wavelengths. The result of this iterative process is shown in the bottom panel of Fig. \ref{resta_agn}. The
AGN fractional contributions to the total 6, 20 and 30~$\mu$m emission are 43, 88 and 85\% respectively. Therefore, we recover the
AGN spectrum of NGC\,2992 using only the generic torus models from \citet{Ramos-Almeida2011a}, and we obtain practically identical AGN fractional 
contributions at 20 and 30~$\mu$m using the two methods described here.

The contribution from SF to the Spizer/IRS spectrum of NGC\,2992 is concentrated on the PAH features at 6.2, 7.7, 8.6 and 
11.3~$\mu$m, as shown in Fig. \ref{resta_agn}. At shorter wavelengths ($\lambda$ $\lesssim$15~$\mu$m) the SB component is stronger than the AGN component, with a $\sim$60-70\%
contribution at 6~$\mu$m. However, at longer wavelengths  ($\lambda$ $\gtrsim$15~$\mu$m) the AGN becomes dominant, reaching $\sim$90\% at 20-25~$\mu$m. At $\lambda$ $>$25~$\mu$m,
the AGN contribution slowly decreases (see also next section), in agreement with the results reported by \citet{mullaney11} for a sample of intermediate luminosity AGN (L$_{2-10keV}\sim$ 10$^{42-44}$ erg~s$^{-1}$), 
whose 6-100~$\mu$m SEDs are best described by a broken power-law that generally peaks between 15-20~$\mu$m, and finally falls strongly at $\lambda$~$\gtrsim$40~$\mu$m.

\section{Physical parameters of the circumnuclear dust emission of the system}
\label{dust}

Dust grains in Seyfert galaxies are heated mainly by SF and nuclear activity, and this radiation is reemited in the IR range. 
The physical properties of large dust grains can be accurately described by a single modified blackbody \citep{Bianchi13},  
and this dust component, heated by the interstellar radiation field, would contain the bulk of the dust mass in a galaxy \citep{Dale12}.
On the other hand, the physical properties of smaller dust grains, i.e. those producing warmer components, are better
described by a blend of multiple modified blackbodies with different temperatures \citep{DraineandLi07}.  

For the sake of simplicity, here we will consider single components to describe the shape of the FIR SEDs of NGC\,2992 and 
NGC\,2993. This is a similar approach to that used by \citet{Perezgarcia01} and \citet{Prieto-acosta03} to reproduce the IR SED 
of Seyfert galaxies %with data from ISO and IRAS, 
as the sum of three components:
1) a warm dust component, produced by dust heated  by SF and/or nuclear activity at T$\sim$120-170 K;
2) a cold component (T$\sim$30-70 K), associated with SF regions; and 3) a very cold 
component (T$\sim$15-25 K), produced by dust heated by the interstellar radiation field. 
The warm component peaks in the MIR range, and both the cold and very cold dust 
components are detected in the FIR range.

%%Dust grains in Seyfert galaxies are heated mainly by SF and nuclear activity, and this radiation is reemited in 
%the IR range. {\bf{According to previous work based on IRAS and ISO data \citep{Perezgarcia01,Prieto-acosta03}, 
%the shape of the SED of Seyfert galaxies can be characterised 
%as the sum of three components, assumed to be modified blackbodies associated to a single temperature and named 
%as warm (T$\sim$120-170 K), cold (T$\sim$30-70 K) and very cold (T$\sim$15-25 K) component, respectively. More 
%recent work has shown that such simple model does not reflect the actual dust grain temperature distribution
%responsible of the infrared emission. Thus, \citep{DraineandLi07} have demonstrated that the physical properties 
%of the warmer components can be accurately described by a  blend of multiple modified blackbodies with different 
%temperatures. 
%On the other hand, \citep{Bianchi13} has shown that the emission in the 
%wavelength range $100~\micron \leq \lambda \leq 500 \micron$ can be well described by a single temperature
%modified blackbody emitted by a population of large dust grains heated by the interstellar radiation field.  
%Furthermore, this dust component contains the bulk of the dust mass in a galaxy \citep{Dale12}.  Under these 
%premises, we will estimate the dust mass content of the  galaxy system based uniquely on the emission at  
%$\lambda \geq 100~\micron$.}}

\begin{table*}
%\scriptsize
\centering
\begin{tabular}{lccccccccc}
\hline
&70$\mu$m&100 $\mu$m&160 $\mu$m&250 $\mu$m&350 $\mu$m&500 $\mu$m& Temperature &M$_{dust}$&SFR\\
&(mJy)&(mJy)&(mJy)&(mJy)&(mJy)&(mJy)& (K) & (10$^{6}$~M$_{\odot}$)&(M$_{\odot}$/yr)\\
\hline
NGC\,2992 (a)&6801$\pm$1075&9013$\pm$1424&6759$\pm$1068&2124$\pm$363&570$\pm$97&40$\pm$7& 29$\pm$1 & 7.6$\pm$1.3&2.5$\pm$0.4\\
NGC\,2992 (b)&1751$\pm$277&2776$\pm$439&3981$\pm$629&1743$\pm$298&1530$\pm$262&316$\pm$54& 21$\pm$1 & 19.6$\pm$2.7&0.7$\pm$0.1\\
NGC\,2993 (a)&10083$\pm$1593&10979$\pm$1735&7153$\pm$1130&1625$\pm$278&488$\pm$83&42$\pm$7& 33$\pm$1 & 4.7$\pm$0.8&3.7$\pm$0.6\\
NGC\,2993 (b)&2055$\pm$325&2902$\pm$459&3623$\pm$572&1666$\pm$285&1266$\pm$216&249$\pm$43& 22$\pm$1 & 14.9$\pm$2.0&0.8$\pm$0.1\\
Arp\,245N    &77$\pm$9&163$\pm$18&394$\pm$44&279$\pm$32&176$\pm$20&31$\pm$4&19$\pm$1&3.2$\pm$1.1& $\sim$0.03\\
\hline
%Arp~245N&48$\pm$8&124$\pm$20&290$\pm$46&204$\pm$35&119$\pm$20&25$\pm$4& 19$\pm$1& 2.3$\pm$1.1\\
%(r$\sim$35\arcsec)\\
%Arp~245~Bridge&63$\pm$10&160$\pm$25&98$\pm$16&72$\pm$12&54$\pm$9&18$\pm$3& 21$\pm$1& 0.8$\pm$1.1\\
%(r$\sim$70\arcsec)\\
%\hline
\end{tabular}						 
\caption{Fluxes of the circumnuclear (a) and disk (b) components of NGC\,2992 and NGC\,2993 described in Section \ref{dust}, total 
fluxes of Arp\,245N and physical properties derived from the SED fits
shown in Fig. \ref{bbodies}. See Fig. \ref{apertures} for further details on the different apertures employed.}

%{\bf{One~FWHM SPIRE~500 aperture fluxes ($\sim$35\arcsec) with the subtraction of the 
%the corresponding galaxy background for NGC~2992 and NGC~2993. One~FWHM SPIRE~500 aperture fluxes with the sky subtraction 
%for Arp~245N and two~FWHM SPIRE~500 aperture fluxes for Arp~245~Bridge ($\sim$70\arcsec) with the sky subtraction. All of them 
%in mJy. For futher details about the apertures see Fig. \ref{apertures} and about the errors see Table \ref{flux_results} and \ref{fluxes}.}}
\label{bbodies_results}
\end{table*}

\begin{figure}
\centering
\includegraphics[width=7.5cm,angle=90]{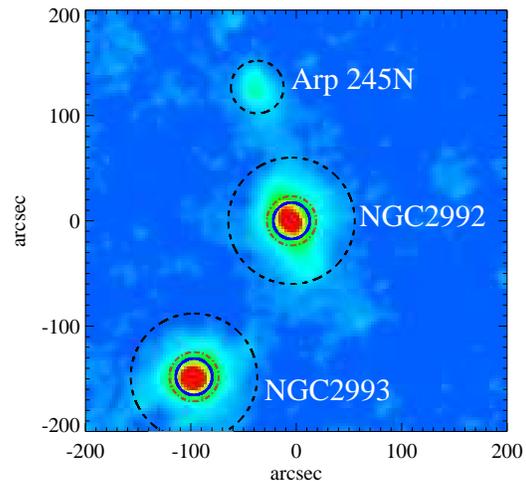}
\caption{Schema of the different apertures used to calculate the circumnuclear and disk emission of NGC\,2992/93, and 
the total flux of Arp\,245N. The apertures are overplotted on the 160 \micron~Heschel/PACS image of the system. 
Blue solid and brown dot-dashed circles correspond to the apertures used to calculate the fluxes of the circumnuclear component and 
the galaxy background that we subtract from the former to get rid of the underlying disk emission. Black dashed circles correspond 
to the apertures used to calculate the total galaxy fluxes reported in Table \ref{flux_results}. The disk emission is calculated by 
subtracting the circumnuclear fluxes from the total emission.}
%%The red solid circle is the aperture total flux of Arp\,245N in an aperture radius of 25\arcsec with sky background subtracted.}
\label{apertures}
\end{figure}

\begin{figure}
\centering
\subfigure[]{\includegraphics[width=6.5cm,angle=90]{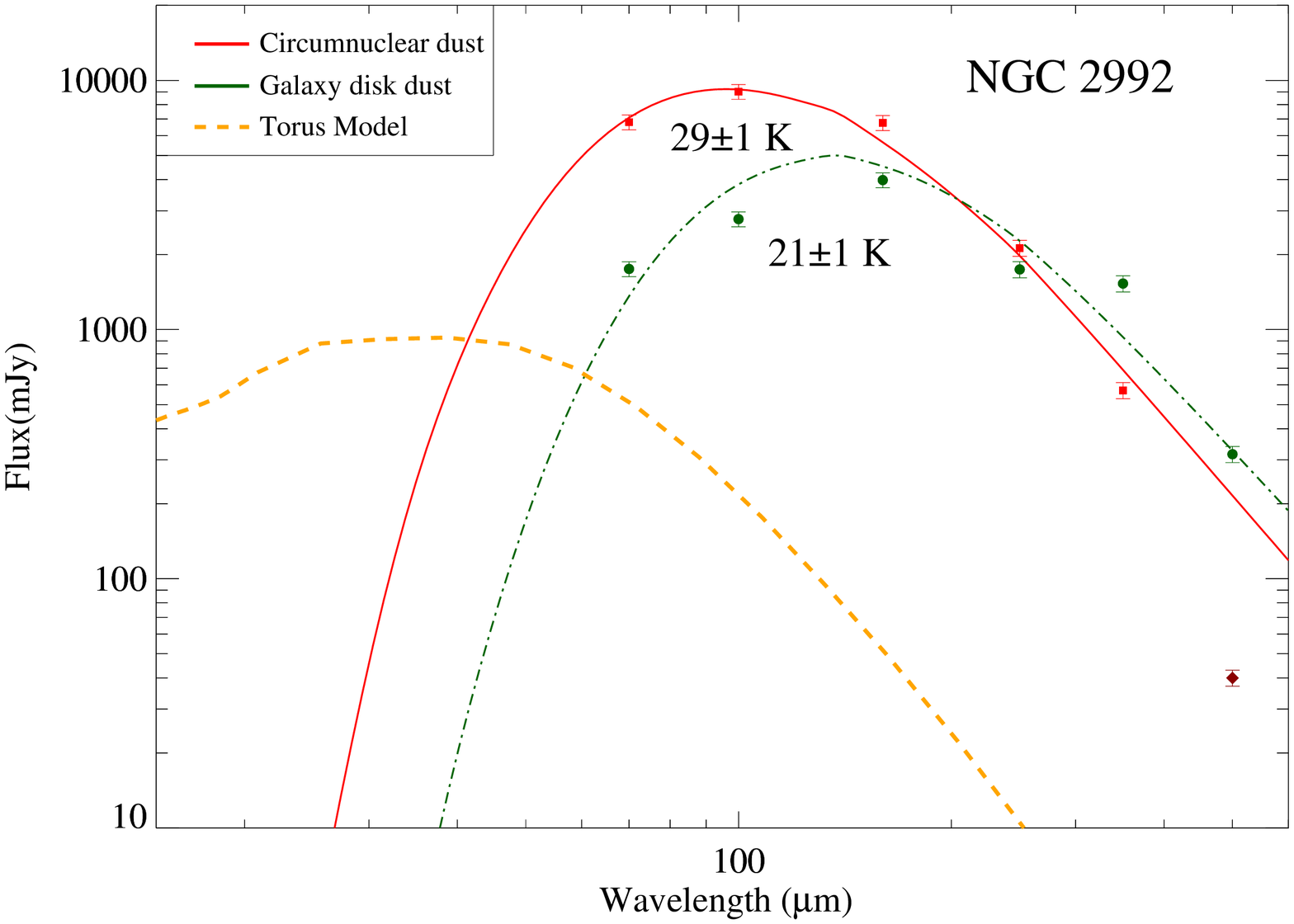}}
\subfigure[]{\includegraphics[width=6.5cm,angle=90]{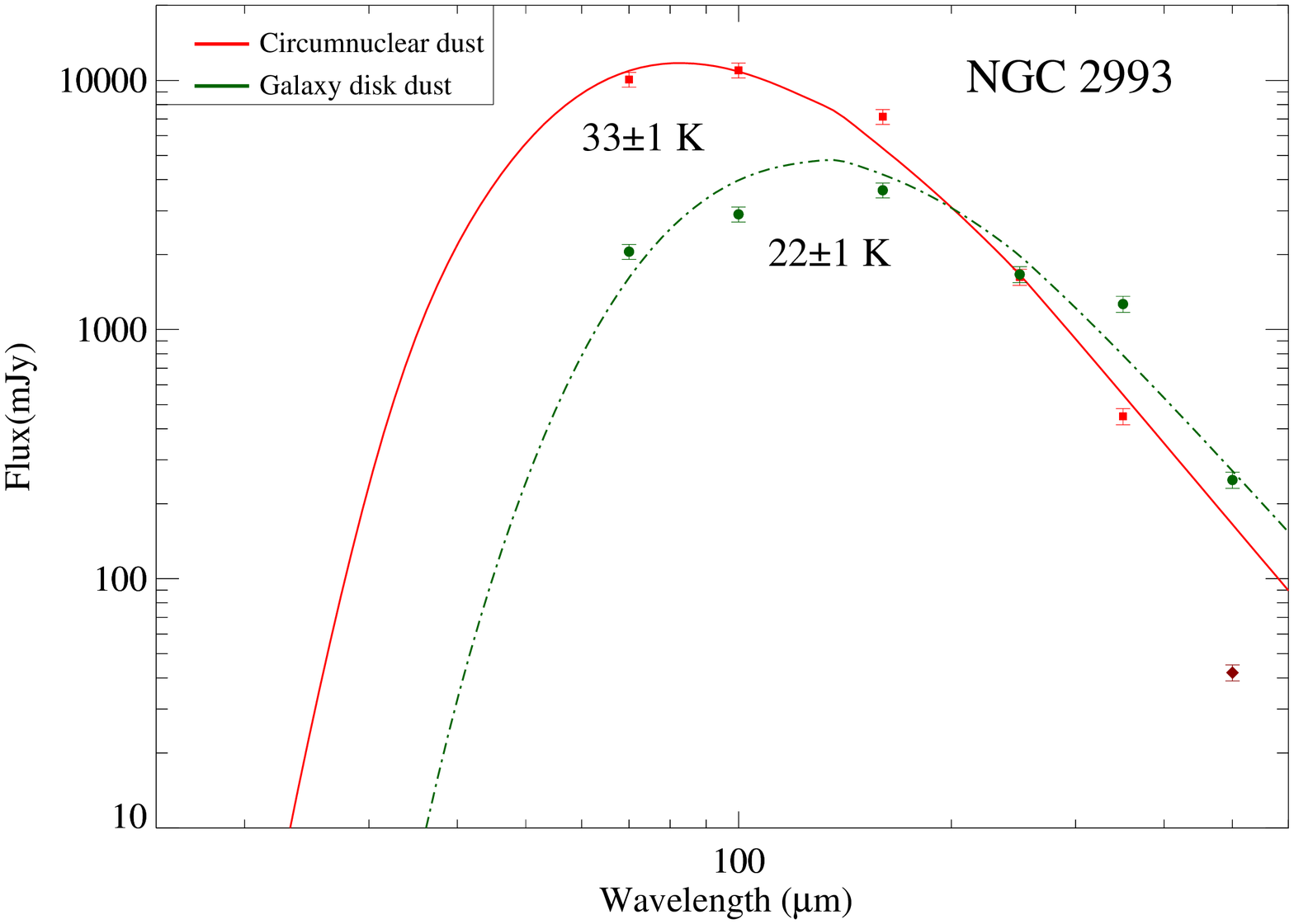}}
\subfigure[]{\includegraphics[width=6.5cm,angle=90]{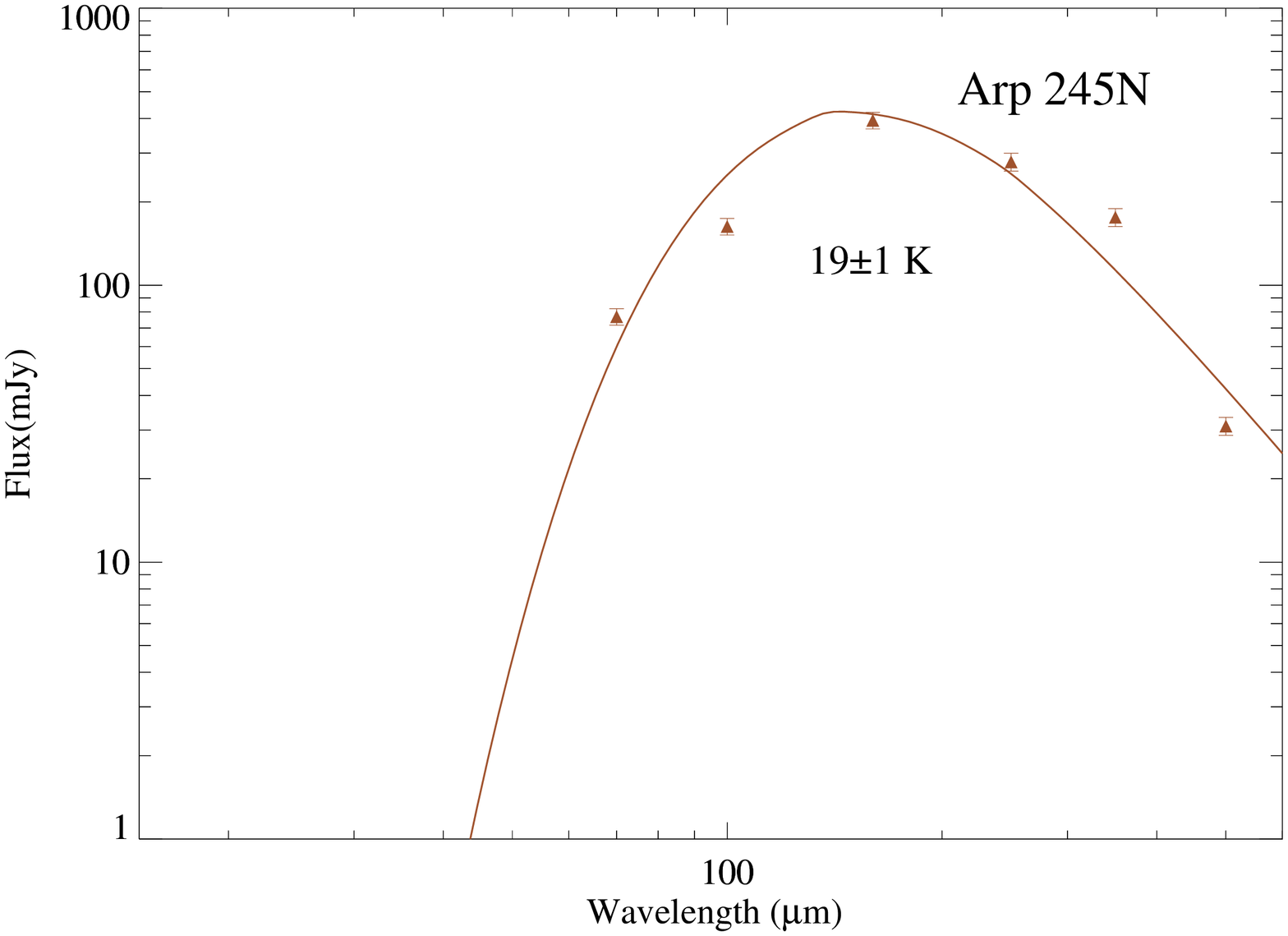}}
\caption{Top and central panels: fits to the circumnuclear and disk SEDs of NGC\,2992/93 (red solid and green dot-dashed lines 
respectively). The MAP torus model fitted in Section \ref{bayesclumpy} is also shown in the case of NGC\,2992 (orange dashed line). 
Bottom panel: SED fit of Arp\,245N (solid brown line). 
Note that in the top and central panels we represent the circumnuclear SPIRE 500~$\mu$m fluxes as dark red diamonds, which are excluded 
from the fits because they are clearly underestimated (see Section \ref{dust}).}
\label{bbodies}
\end{figure}

In the case of the galaxies NGC\,2992/93, we consider two different regions to separate the circumnuclear emission  
from the diffuse and more extended disk emission, as we did in \citet{Ramos-Almeida2011c} for the Seyfert 2 galaxy NGC\,3081.
The circumnuclear region is delimited by a circular aperture of 
$\sim$35\arcsec~diameter ($\sim$6 kpc), coincident with the largest FWHM of the Herschel data (i.e. 
SPIRE 500 \micron; blue solid circles in Fig. \ref{apertures}). We first extracted Herschel/PACS and SPIRE 
fluxes within such aperture and then subtracted the corresponding galaxy background measured in an adjacent 
annulus (brown dot-dashed circles in Fig. \ref{apertures}).
In the case of the active galaxy NGC\,2992 we also subtracted the torus model emission (described in Section \ref{bayesclumpy}), 
which is the main contributor to the warm dust component in this galaxy. The circumnuclear fluxes are represented as red squares
in Fig. \ref{bbodies} and reported in Table \ref{bbodies_results}. 

The disk emission is computed as the difference between the total fluxes reported in Table \ref{flux_results} (calculated
in apertures of 120\arcsec~diameter; black dashed circles in Fig. \ref{apertures}) 
minus the circumnuclear fluxes. These values are plotted as green circles in Fig. \ref{bbodies} and reported 
in Table \ref{bbodies_results}.

In order to quantify the temperatures and dust masses of the two components described above
we used the following relation:
$M_{dust}=D^2 f_{\nu}/\kappa_{\nu}B(T)$ \citep{Hildebrand83,Bianchi13} to fit the FIR SEDs. $D$ is the luminosity distance to the galaxy, 
$f_{\nu}$ the flux density, $\kappa_{\nu}$ the absorption opacity of the dust at frecuency $\nu$\footnote{The average
absorption cross section is available at 
http://www.astro.princeton.edu/$\sim$draine/dust/dustmix.html. In this paper
we use the $R_V=3.1$ MW dust model. See \citet{LiandDraine01} for further details.},
and $B(T)$ the Planck function evaluated at temperature T.

The fits to the circumnuclear and disk FIR SEDs of NGC\,2992/93 are shown in the top and central panels 
of Fig. \ref{bbodies}\footnote{Note that we excluded the circumnuclear SPIRE 500 \micron~fluxes from the fits, as they 
are underestimated because the circumnuclear aperture does not include an important part of the unresolved flux in this 
band.}. The best fits correspond to dust temperatures of
29$\pm$1~K and 33$\pm$1~K respectively (see Table \ref{bbodies_results}). These temperatures explain the similarity 
between the total MIR-to-FIR
SEDs of the two spiral galaxies (see Fig. \ref{SED}). On the other hand, the SEDs of the disks can be described by a dust component of  
21$\pm$1~K and 22$\pm$1~K respectively (see Table \ref{bbodies_results}). These temperatures
coincide with the lowest values reported by \citet{Skibba2011} for the nearby galaxies in the KINGFISH Herschel survey 
(T$\sim$20-35~K). They are also 
consistent with the values measured for the coldest dust components of Seyfert galaxies 
(e.g. \citealt{Radovich99,Ramos-Almeida2011c,Bendo10}) using Herschel data as in this study.

%{\textcolor{blue}{{\bf{We have also excluded from the fits the
%flux at 70~$\mu$ since it could be attributed to a dust grain population different than the large grains, however we
%found that our single-temperature modified blackbody models can reproduce quite well the measured fluxes (see Fig. \ref{bbodies}),
%indicating that the emission likely corresponds to the same dust grain population.}}}}%

%derived by fitting
%a modified blackbody with $\beta=2$. Most of studies suggest values of $\beta$ between 1 and 2, with some dependence on the
%grain size and the temperature (e.g. \citealt{LiandDraine01}). Typically, these studies favour $\beta\sim$1 for small carbonaceous grains, which radiates most of their energy in the MIR range,
%and $\beta\sim$1.5-2 for big silicates grains, which provide the lower temperature and radiate most of their energy at FIR and submillimetre wavelengths. Since, here 
%we fitted FIR data, we used modified
%blackbodies of emissivity $\beta$=2 (i.e. greybodies).}}

The dust masses that we measure for the two components are reported in Table \ref{bbodies_results}. We obtain
similar values for the two spiral galaxies: $\sim$5-$8\times10^6$~M$_{\odot}$ and  $\sim$15-$20\times10^{6}$~M$_{\odot}$ 
for the circumnuclear and galaxy disk components respectively, with those of NGC\,2992 being slightly larger. 
These dust masses are within the lowest values reported by \citet{daCunha10} for a sample of star-forming galaxies, 
using the M$_{dust}$-star formation rate (SFR)
relation ($M_{d}\sim$10-100$\times10^{6}$~M$_{\odot}$ for the galaxies with the highest S/N of the
sample).

We also measured total fluxes for the dwarf galaxy Arp\,245N in an aperture radius of 25\arcsec, after subtracting the 
sky background (see Table \ref{bbodies_results} and Fig. \ref{apertures})\footnote{These fluxes are same as those reported
in Table \ref{flux_results}.}. 
These fluxes are plotted as brown triangles in the bottom panel of Fig. \ref{bbodies}, and fitted using a temperature  
of 19$\pm$1~K and a total dust mass of $M_{dust} = (3.2 \pm 1.1) \times 10^6 M_{\odot}$. 
This low temperature is compatible with dust heated by the interstellar radiation field and the relatively 
weak MIR emission of the galaxy in the range $\sim$24--100 \micron~(see Fig. \ref{SED}) would be indicating that
a significant part of the old stellar population and the interstellar medium of NGC\,2992 and NGC\,2993
have been stripped from the galaxies during the interaction process undergone by the system.

%AGN contribution, although we subtracted the torus emission from the circumnuclear dust component. In fact, the SFRs 
%that we measured for the two spiral galaxies are very similar considering the errors (see Table \ref{bbodies_results}).

%Using the relation $M_{d}~=~7.9\times10^{-5}~(T_{K}/40)^{-6}~L_{IR}/L_{\odot}$ from \citet{klaas93}, where T$_K$ corresponds 
%to the temperatures of each greybody, 

Finally, we estimated SFRs for the three galaxies in the system using the 70 \micron~fluxes reported
in Table \ref{bbodies_results} and equation 14 in \citet{Rieke09}. We decided to use the 70 \micron~fluxes because the 24 \micron~flux
is dominated by the AGN contribution in the case of NGC\,2992. However, we checked that we obtain similar SFR estimations 
for NGC\,2993 and Arp\,245N when using 24 and 100 \micron~fluxes. The SFRs 
that we measure for the two spiral galaxies are very similar considering the errors (see Table \ref{bbodies_results}). 
For Arp\,245N we derive a small SFR ($\sim$0.03 $M_\odot /yr$), consistent with the low temperature of the dust 
and the weak MIR emission. 

\section{Discussion}
\label{Discussion}

\subsection{NGC 2992}
\label{disc_2992}

NGC\,2992 contains a Seyfert 1.9 nucleus, although it has changed its type between Seyfert 1.5 and 2 and 
it has also exhibited huge variations accross all the spectrum \citep{Trippe08}. The 2-10 keV X-ray luminosity 
dramatically decreased between 1978 and 1994 (a factor of $\sim$16; see Table \ref{x_ray_fluxes}) and then increased by roughly the 
same amount from 1994 to 2005 \citep{Gilli00,Yaqoob07,Brightman11}. 
Besides, \citet{Glass97} reported variability in the NIR, with the source fading from 1978 to 1996, except for 
a strong outburst in 1988. This IR variability mirrors that detected in the X-ray regime, with the corresponding delay due to the 
different scales that each emission is probing \citep{Clavel89,Barvainis92,Honig11}. 
\citet{Gilli00} explained the extreme variability of NGC\,2992 as caused by a retriggered AGN, and in particular, by different stages 
of the rebuilding of the accretion disk, which the latter authors estimated to range between 1 and 5 years. 

According to the new observations reported here, the nuclear MIR spectrum of NGC\,2992 has not changed either in flux or shape
from 2007 to 2014. This is in agreement with the scenario proposed by \citet{Gilli00}, in which the rebuilding of the 
accretion disk would have been finished in 2005, first stopping the X-ray variability and finally, the IR variability.
% Considering that the periods of nuclear activity are restricted to a few Myr, depending on the AGN luminosity, 
%and likely intermitent, 
Monitoring campaigns in the X-ray and the optical/IR are then key to constrain the relative sizes of
the AGN internal structures as well as to understand the physics of nuclear activity.

%Before the BeppoSAX observations of NGC\,2992 (1997--1998; see Table \ref{}), all evidence suggested that the AGN in the center was off, 
%but the X-ray flux measured by BeppoSAX increased by a factor of about 12 \citep{Gilli00}, suggesting AGN retriggering.
%{\bf{Que paso despues...
%NGC\,2992 was observed by Suzaku  in 2005 and the data showed that the intensities of 
%RXTE IN HTE PERIORD 200Panel a shows the orientation of the Spitzer/IRS slits.5 March to 2006 January.}}

%Although the observed variations of the SED of NGC\,2992 could be due to a variable ionizing continuum, as suggested  by \citet{Trippe08}, 
%it is possible that this variability is associated with the obscuring material instead (e.g. a clumpy torus). This would explain 
%not only IR variability, but also the changes in Seyfert type. 
In Section \ref{bayesclumpy} we derived a column 
density of N$_H\sim10^{24}$~cm$^{-2}$ from the fit of the nuclear SED with clumpy torus models. This column density is much higher than the values 
derived from X-ray measurements, as e.g. the N$_H\sim10^{22}$ cm$^{-2}$ reported by \citet{Gilli00}
or the N$_H=8$~x~$10^{21}$ cm$^{-2}$ reported by \citet{Yaqoob07}. However, \citet{Weaver96} measured a narrow and prominent Fe K$\alpha$ line with a large equivalent width ($\sim 500~$eV) from X-ray data taken with the Advanced Satellite for Cosmology
and Astrophysics (ASCA). The strong Fe K$\alpha$ line and the Compton-reflection 
component inferered by \citet{Weaver96} 
require the existence of cold dense gas with a column of density of
N$_H\sim10^{23}$-$10^{25}$~cm$^{-2}$, which comprises the value derived from our fit.
%However, \citet{Weaver96} measured a narrow and prominent 
%Fe K$\alpha$ line with a large equivalent width ($\sim 500~$eV) from X-ray data taken with the Advanced Satellite for Cosmology
%and Astrophysics (ASCA). The strong Fe K$\alpha$ line is probably a relativistic line emitted by the accretion disk \citep{Shu10}, which
%could explain the X-ray variability. 
Moreover, \citet{Weaver96} associated the lag 
in the response of these X-ray features to changes in continuum flux, from which they estimated the reprocessor size to be 
$\sim$~3~pc. This is roughly in agreement with the torus size derived from our fit (R$_o$=1.5$\pm$0.5 pc). Therefore,
a clumpy torus with the properties derived from the fit perfomed here is in agreement with the scenario proposed by
\citet{Weaver96}. On the other hand, the intermediate covering factor of the torus that we infer from the fit presented in Section \ref{bayesclumpy}
could explain the changes in Seyfert type experienced by NGC\,2992 \citep{Trippe08}.

In Section \ref{decomposition} we used clumpy torus models and a set of SB templates to decompose the
Spitzer/IRS spectrum of NGC\,2992, which probes the central kpc of the galaxy. We found an important contribution 
from SF at short MIR wavelengths ($\lambda\sim$6-15~$\mu$m; 60-70\%), which is concentrated on the PAH 
features at 6.2, 7.7, 8.6 and 11.3~$\mu$m. On the other hand, at $\lambda\sim$15-30~$\mu$m, the AGN dominates 
the Spitzer/IRS spectrum, reaching 90\% at 20-25~$\mu$m. A similar, but more simplistic approach was taken by \citet{Deo09}
using Spitzer/IRS data in mapping-mode\footnote{In this work we are using the Spitzer/IRS spectrum of NGC\,2992 in 
staring-mode.}. They subtracted the average starburst galaxy spectrum from \citet{Brandl06} from the Spitzer/IRS 
spectrum after scaling it, aiming to completely remove the 11.3 and 17~$\mu$m~PAH features from the residual. 
By using this method, they found that the contribution from the SB increases with wavelength, as opposed to 
what we find using our spectral decomposition technique. We also tried a fit using clumpy torus models and 
the \citet{Brandl06} SB template, but the result did not sucessfully reproduce the Spitzer/IRS spectrum. In 
addition, from the analysis of the IR photometry performed in Section \ref{recover} we know that the AGN emission, in the 
scales probed by Spitzer, becomes dominant at 20-25~$\mu$m, in agreement with the results derived from spectroscopic data. 
Finally, using the high angular resolution imaging and spectroscopic data of NGC\,2992 presented here, we have shown that 
SF is either supressed or diluted by the strong AGN radiation field in the inner $\sim$50 pc of the galaxy (See Section \ref{spectroscopy}),
as shown by the lack of PAH features in the nuclear spectrum.

\subsection{The Arp 245 interacting system}
\label{disk_245}

As previously mentioned, the interacting system Arp\,245 is formed by the two spiral galaxies NGC\,2992/93, 
the dwarf galaxy Arp\,245N and two major tidal features. These features consist on two bridges connecting the three 
galaxies (see Fig. \ref{ngc2992__all}). The fact that the tidal features are well-developed indicates that the system is 
seen close after its first encounter (see \citealt{duc00} and references therein). Furthermore, the northern bridge linking 
NGC\,2992 and  Arp\,245N is relatively bright and dense, whereas that associated with NGC\,2993 appears like a 
weaker large open ring. The prominence of the two bridges suggests that the two spiral galaxies are experiencing  
prograde encounters (galaxy spins in the same sense as the flyby; \citealt{Toomre72}). 
Nonetheless, the bridges extend $\sim$16 and 27~kpc respectively, which are rather modest extents compared 
with the long 100~kpc 
antennae observed in the prototypical interacting galaxy pair NGC~4038/39. However, these lengths are expected
for an interacting system in an early phase such as Arp\,245, with the bridges still developing.
This is confirmed by the numerical modeling performed by \citet{duc00}, which predicts that the first galaxy 
encounter happened $\sim$100 Myr ago. 

According to \citet{Tadhunter11} and \citet{Ramos-Almeida2011b}, three main stages can be defined in a galaxy merger 
sequence: {\it i)} 
pre-coalescence,{\it ii)} coalescence, and {\it iii)} post-coalescence. In the pre-coalescence phase the 
two nuclei are observed after the first passage, and immediately before ($\sim$100 Myr) 
the coalescence of the two nuclei. The peak of AGN activity is expected 
during coalescence, although during the pre-coalescence most objects tend to also exhibit AGN or SB 
activity associated with the gas infall produced by the tidal forces at play. 
This is indeed the case of the galaxy pair NGC\,2992/93:  NGC\,2992 is known  to host a 
Seyfert nucleus, and all the galaxies show circumnuclear SF, as revealed by 
their FIR luminosities (see Section \ref{large}) and the H$\alpha$ imaging presented in
\citet{duc00}. 

A more recent study on interacting systems is presented in \citet{Lanz13}, where the authors measured and modelled 
the galaxy SEDs from the ultraviolet to the FIR, also using Spitzer and Herschel data to cover the IR range. From this 
modelling they derived temperatures, dust masses and SFRs for the galaxies in their sample. In addition, \citet{Lanz13}
classify the galaxies in different interaction stages attending to the galaxy morphologies, using a similar classification scheme 
as in \citet{Dopita02}. They divide the sample in four categories: 1) non-interacting 
galaxies; 2) galaxies in a weak integrating system, which are close but show minimal morphological distortion;
3) moderately interacting galaxies showing strong sings of morphological disturbance such as tidal tails; and
4) strongly interacting galaxies in a more evolved stage of the interaction.

As explained above, the Arp\,245 system would be in the pre-coalescence stage of the interaction, which would correspond to 
stages 3/4 of the classification employed by \citet{Lanz13}. For these stages they report average dust masses of 
$\sim$1--4$\times10^7$ M$_{\odot}$, dust temperatures of $\sim$20--23 K, and SFRs of $\sim$0.6--7.8~${M}_\odot$/yr.
These values are consistent with the measurements reported in Table \ref{bbodies_results} for the disk components of 
NGC\,2992/93, being more similar to the stage 3 values, i.e., moderately integrating galaxies showing strong 
signs of morphological disturbance. 

Using the determined dust masses for the spiral galaxies, we derived 
gas-to-dust ratios\footnote{The HI mass has been taken from \citet{duc00}. We note that we calculate the gas-to-dust ratios using
atomic gas masses, but it is possible that a substantial fraction of the gas mass is in molecular form.} of 70 and 50 for NGC\,2992 and NGC\,2993 respectively. 
These ratios are comparable to those measured in other nearby galaxies, including active and non-active galaxies. As an example, 
see the results obtained for the Spitzer IR Nearby Galaxies Survey (SINGS; \citealt{Draine07}).

%{\bf{We have also calculated dust masses for the interacting galaxies by fitting their SEDs using the dust models of \citet{LiandDraine01}.}} 

A remarkable feature is the similarity of the MIR-to-FIR (beyond 10~$\mu$m) emission coming from 
the two spiral galaxies (see Figure \ref{SED}). They are closely matched in luminosity and spectral shape. 
We can compare the SFRs derived from our FIR data with those 
obtained by \citet{duc00} from H$\alpha$ imaging. 
As expected, the largest contribution to the SFR comes from the circumnuclear region  
( ratio 4-5:1 relative to the disk). Adding the two contributions we obtain 3.2 and 4.5 M$_{\odot}$~yr$^{-1}$
for NGC\,2992 and NGC\,2993, respectively. \citet{duc00} reported 
H$\alpha$ luminosities of 1.7 and 2.8$\times 10^{41}~\mathrm{erg~s^{-1}}$ for the two spirals, 
which can be transformed to SFRs, resulting in 0.7 and 1.14~$\mathrm{M}_\odot~yr^{-1}$ for NGC\,2992 and NGC\,2993
respectively. We corrected these figures for extinction using E(B-V)=0.84 and 0.7 for NGC 2992/93, measured from 
the recombination lines \citep{Durret88}. 
The extinction-corrected values are 9.1 and 9.8~$\mathrm{M}_\odot~yr^{-1}$ for NGC\,2992 and 
NGC\,2993 respectively, which are larger than the values that we derive here. However, we note that the average 
extinction correction that we are using may not be adequate, 
since the integrated H$\alpha$ emission comes from regions with different levels of obscuration.

%This SFR derived from the FIR emission agrees with that derived by \citet{duc00} from the 
%H$\alpha$ map, although our values are much lower than their extinction-corrected estimations.  
%Nevertheless, we cannot disregard the discrepancy due to the different apertures selected for the 
%calculations. The gas-to-dust mass ratio is 360, which might indicate that this region is a 
%HI reservoir \citep{duc00}, where the dust content is relatively low, as well as the SF. 

All the tidal features observed in the optical images of the Arp\,245 system \citep{duc00} have counterparts in the 
MIR-to-FIR observations presented here (see Fig. \ref{ngc2992__all}), although they show different properties.   
The galaxy Arp\,245N, that is clearly detected in the four Spitzer/IRAC bands (from 3.6 to 8~$\mu$m), becomes much
fainter at 24 and 70~$\mu$m, and arises again at wavelengths beyond 100~$\mu$m. This bimodal emission fits with 
the hypothesis made
by \citet{duc00}, who proposed this source to be formed by an old stellar population tidally stripped from 
NGC\,2992, plus a minor contribution from young stars formed in-situ after the interaction. The latter is
spectroscopically corroborated by the presence of H$\alpha$ and H$\beta$ emission. The EW of H$\beta$ emission 
line indicates that the SB started less than 10 Myr ago \citep{duc00}. 
Using the PACS~70~$\mu$m flux of Arp\,245N, we derived 
a relatively low SFR per unit area ($\Sigma$SFR): Log $\Sigma$SFR$\sim$ -3.3$\mathrm{M}_\odot~yr^{-1}~kpc^{-2}$,
which is considerably larger than the extinction-corrected value reported by \citet{duc00},
of -2.5 $\mathrm{M}_\odot~yr^{-1}~kpc^{-2}$, obtained from the H$\alpha$ luminosity.
This difference could be due to an overestimation of the extinction correction applied by \citet{duc00}, who
measured Log $\Sigma$SFR$\sim$ -3.1$\mathrm{M}_\odot~yr^{-1}~kpc^{-2}$ before correcting for extinction.
The gas-to-dust mass ratio that we measured for Arp\,245N is $\sim$280, which might indicate that this region is a 
HI reservoir, where the dust content is relatively low, as well as the SF.

%\citet{Duc00}  
%broadly discussed the nature of this tidal tail as a Tidal Dwarf Galaxy, and they concluded that  it is in the way to 
%reach such conditions to be classified as a bounded system. 
The Arp\,245 Bridge is only detected at wavelengths longer than 160~$\mu$m 
(see Fig. \ref{ngc2992__all}), indicating the presence of very cold dust. 
%{\textcolor{blue}{We calculated a 
%lower SFR per unit area from Arp\,245 Bridge (log(SFR/A)=$-$(1.64$\pm$0.34)).}} 
There is no evidence for dust heated by SF  
activity in this feature, which is corroborated by the lack of ionized gas in the H$\alpha$ images  
\citep{duc00}. The lack of recent SF in the bridge would indicate that the conditions are not adequate, 
i.e., the HI column density could be below the threshold required for SF to take place. 
%We measured a 
%gas-to-dust ratio of $\sim530$ in the bridge, which is higher than those measured in e.g. galaxy disks. 
% {\bf{The big gas-to-dust ratio would be consistent with
%what could be expected for low-metallicity gas stripped from the outer radii of galaxies \citep{munoz-mateos09}.}}

Summarizing, the MIR-to-FIR maps presented in this work are in good agreement with the system being in an early 
phase of the interaction between the galaxy pair NGC\,2992/93 ($\sim$100 Myr after the first encounter).
The MIR-to-FIR luminosities 
indicate that both spirals are relatively bright IR galaxies, with the SF activity mostly concentrated in the 
circumnuclear regions. On the other hand, the tidal features have not reached the conditions 
to be active star forming sites. 

IR studies of interacting systems are important to advance in our
understanding of the evolution of the gas/dust properties of such systems as the 
interaction evolves. Our work contributes to this understanding with the analysis of two spiral galaxies in clear interaction, 
one of them an AGN, and a dwarf galaxy likely stripped from NGC\,2992. The different dust properties of 
the dwarf galaxy and the two spirals provide one more piece of information about how star formation and nuclear activity are
triggered in galaxy interactions, but similar studies of other interacting systems in different stages of the interaction, 
including galaxies of different masses and types, are fundamental to derive general conclusions.

%We derived a
%temperature of the cold dust of $\sim$22~K, which is consistent with an intermediate temperature between
%the average temperature reported by \citet{Lanz13} for the stage 3 ($\sim$20~K)
%and 4 ($\sim$23~K). We also derived the dust mass of both spiral galaxies, again in agreement with the average mass dust
%reported by \citep{Lanz13} between stage 3 ($\sim$1$\times10^7$ M$_{\odot}$) and 4 ($\sim$4$\times10^7$ M$_{\odot}$).}}

\section{Conclusions}
\label{Conclusions}

We have presented Spitzer and Herschel IR imaging of the interacting system Arp\,245, and high angular resolution 
IR imaging and MIR spectroscopic observations of 
the Seyfert 1.9 galaxy NGC\,2992. For NGC\,2992, we have used different methods to recover the nuclear emission 
from the Spitzer and Herschel data, and compared it with the ground-based IR observations of this galaxy. We
have also studied in detail the circumnuclear and disk emission of the Arp\,245 system, and reproduced the FIR SEDs 
of these two regions with dust models, from which we derived dust temperatures
and masses. Our major results are as follows: \\ 

\textbullet \ \  The ground-based 11.2 $\mu$m image of NGC\,2992 has an angular resolution of 0.32\arcsec~(55 pc) and 
reveals faint extended emission along PA$\sim$30$^\circ$ and out to $\sim$3~kpc. 
The orientation of this extended emission coincides with the semi-major axis of the galaxy. 

\textbullet \ \ The GTC/CC spectrum of the faint MIR extended emission clearly shows the 11.3~$\mu$m PAH feature and the [S~IV]$\lambda$10.5$\mu$m emission line
once we remove the AGN contribution. Therefore, we conclude that this extended emission is produced, at least in part, 
by dust heated by star formation. Moreover, by comparing the extended and nuclear spectra of the galaxy we conclude that either the PAH 
features have been destroyed in the inner $\sim$50 pc of NGC\,2992, or are diluted by the strong AGN continuum. 
 
\textbullet \ \ The GTC/CC and Gemini/MICHELLE nuclear spectra of NGC\,2992, which probe the inner $\sim$50 pc of the
galaxy, are identical in spite of the time difference between the observations ($\sim$7 years) and the different 
slit orientations. These spectra show [S~IV]$\lambda$10.5$\mu$m emission and no PAH features. This similarity indicates that 
the X-ray and IR variability previously observed in this galaxy may have stopped after 2007.

\textbullet \ \  We modelled the nuclear IR SED of NGC\,2992 with clumpy torus models, and derived an AGN bolometric luminosity 
of L$_{bol}^{AGN}$=5.8$\times10^{43}$ erg~s$^{-1}$, 
consistent with the value estimated from X-ray data: L$_{bol}^{X-ray}$=3.2$\times10^{43}$ erg~s$^{-1}$.
We infer a small torus radius of $\sim$1.2 pc from the fit, a torus mass of
M$_{torus}$=9$\times10^4$M$_{\odot}$ and a column density of N$_H$=3.4$\times10^{24}$ cm$^{-2}$. The latter 
value is consistent with the range reported from X-ray observations of the inner $\sim$3 pc of the galaxy.

\textbullet T\ \ We calculated nuclear fluxes for NGC\,2992 using the Spitzer MIR and Herschel FIR images and 
different methods, and we found that we can only recover the nuclear fluxes obtained from high angular resolution
data at 20-25 $\mu$m, where the AGN emission dominates. 

\textbullet \ \ We decomposed the 5-30~$\mu$m Spitzer/IRS spectrum, which probes the inner 630 pc of NGC\,2992, in AGN and SB 
components. We found that the SB component dominates the MIR emission at $\lambda\lesssim 15 \mu$m,
with $\sim$60-70\% contribution at 6 $\mu$m. At $\lambda\gtrsim 15 \mu$m, the AGN component dominates, reaching 
90\% at 20 $\mu$m, and decreasing rapidly at $\lambda >$ 30~$\mu$m. 

\textbullet \ \ The scaled AGN template derived from the spectral decomposition of the Spitzer spectrum 
agrees well in flux and shape with the GTC/CC nuclear
MIR spectrum within the errors, proving the reliability of this method for estimating the SB and AGN contribution to the MIR emission.

\textbullet \ \  The MIR-to-FIR total SEDs of the interacting galaxies NGC\,2992/93 are practically identical in 
shape and flux. This similarity is likely related to the presence of intense SF in both galaxies, which is 
heating the dust at similar temperatures.  

\textbullet \ \ We reproduced the FIR emission of the different components of the Arp\,245 
system using dust models and measured practically the same dust masses, temperatures and SFRs 
for NGC\,2992/93. These measurements are very similar to those reported for non-active interacting systems 
in the first stages of the interaction.

\textbullet \ \ The MIR-to-FIR maps and cold dust properties presented here are consistent with the Arp\,245 system being in an early 
stage of the interaction between the galaxy pair NGC\,2992/93, with the SF activity mostly concentrated in their 
circumnuclear regions. On the other hand, the tidal features do not seem to have reached the conditions 
to be active star forming sites.

\section*{Acknowledgments}

IGB ackowledges financial support from the Instituto de Astrof\'isica de Canarias through
Fundaci\'on La Caixa. This research was partly supported by a Marie Curie Intra European Fellowship
within the 7th European Community Framework Programme (PIEF-GA-2012-327934).
CRA and IGB ackowledge financial support from the Spanish Ministry of Science and Innovation (MICINN) through project
PN AYA2013-47742-C4-2-P.  AAH ackowledges support from grant AYA2012-31447.
P.E. acknowledges support from the Spanish Plan Nacional de Astronom\'ia y Astrof\'isica under
grant AYA2012-31277. OGM ackowledges support from grant AYA2012-39168-C03-01. TDS was supported 
by ALMA-CONICYT grant number 31130005.

This work is based on observations made with the Gran Telescopio CANARIAS (GTC), installed in the Spanish
Observatorio del Roque de los Muchachos of the Instituto de Astrof\' isica de Canarias, in the island of La Palma.

This research has made use of the NASA/IPAC Extragalactic Database (NED) which is
operated by the Jet Propulsion Laboratory, California Institute of Technology, under 
contract with the National Aeronautics and Space Administration.

Based on observations obtained at the Gemini Observatory, which is operated by the Association of Universities for
Research in Astronomy, Inc., under a cooperative agreement with the NSF on behalf of the Gemini partnership: the
National Science Fundation (United States), the Science and Technology Facilities Council (United Kingdom), the
National Research Council (Canada), CONICYT (Chile), the Australian Research Council (Australia), Minist\'erio da
Ci\^encia e Tecnologia (Brazil) and Ministerio de Ciencia, Tecnologia e Innovaci\'on Productiva (Argentina).

Based on observations made with the NASA/ESA Hubble Space Telescope, obtained from the data archive
at the Space TElescope Science Institute. STScI is operated by the Association of Universities for Research in
Astronomy, Inc. under NASA contract NAS 5-26555.

Based on observation made with the Spitzer Space Telescope, which is operated by the Jet Propulsion Laboratory,
California Institute of Technology, under NASA contract 1407.

Based on observation made with the Herschel Observatory, which is an ESA space observatory with science instruments 
provided by European-led Principal 
Investigator consortia and with important participation from NASA. PACS has been developed by
a consortium of institutes led by MPE (Germany) and including UVIE (Austria); KU Leuven, CSL, 
IMEC (Belgium); CEA, LAM (France); MPIA (Germany); INAF-IFSI/OAA/OAP/OAT, LENS, SISSA (Italy);
IAC (Spain). This development has been supported by the funding agencies BMVIT (Austria), 
ESA-PRODEX (Belgium), CEA/CNES (France), DLR (Germany), ASI/INAF (Italy) and CICYT/MCYT (Spain).
SPIRE has been developed by a consortium of institutes led by Cardiff Univ. (UK) and including Univ.
Lethbridge (Canada); NAOC (China); CEA, LAM (France); IFSI, Univ. Padua (Italy); IAC (Spain); Stockholm
Observatory (Sweden); Imperial College London, RAL, UCL-MSSL, UKATC, Univ. Sussex (UK); and Caltech,
JPL, NHSC, Univ. Colorado (USA). This development has been supported by national funding agencies:
CSA (Canada); NAOC (China); CEA, CNES, CNRS (France); ASI (Italy); MCINN (Spain); SNSB (Sweden); STFC,
UKSA (UK); and NASA (USA).

The authors are extremely grateful to the GTC staff for their constant and enthusiastic support, specially 
to Carlos \'Alvarez. We finally acknowledge useful comments from the anonymous referee.

\label{lastpage}

\end{document}